\documentclass[11pt]{article}
\usepackage[numbers]{natbib}
\usepackage{amssymb}
\usepackage{amsmath}
\usepackage{commath}
\usepackage{stmaryrd}
\usepackage{bm}
\usepackage{siunitx}
\usepackage{subfig}
\usepackage{algorithm}
\usepackage{algorithmicx}
\usepackage{algpseudocode}
\usepackage{graphicx}
\usepackage{hyperref}
\usepackage{amsthm}
\usepackage{booktabs}

\usepackage{authblk}

\DeclareGraphicsExtensions{.png}
\newcommand{\state}[3]{\underline{#1}[#2]\langle#3\rangle}
\newtheorem*{rmk}{Remark}

\title{A Nonlocal Model for Dislocations with Embedded  Discontinuity Peridynamics}

\author[a]{Teng Zhao\thanks{zhaoteng.me@sjtu.edu.cn}} 
\author[a]{Yongxing Shen\thanks{Corresponding author: yongxing.shen@sjtu.edu.cn}}
\affil[a]{University of Michigan - Shanghai Jiao Tong University Joint Institute, Shanghai Jiao Tong University, Shanghai, 200240, China}

\begin{document}

\renewcommand*{\Affilfont}{\small\it} 
\renewcommand\Authands{ and } 
\date{}
\maketitle

\begin{abstract}
\noindent We develop a novel nonlocal model of dislocations based on the framework of peridynamics.
By embedding interior discontinuities into the nonlocal constitutive law, the displacement jump in the Volterra dislocation model is reproduced, intrinsic singularities in classical elasticity are regularized, and the surface effect in previous peridynamics models is avoided. The extended embedded discontinuity peridynamics overcomes unphysical dissipation in treating discontinuity and is still easy to be solved with the particle-based meshless method.
The properties of the proposed dislocation model are compared with classical elasticity solutions under the case of an edge dislocation, double edge dislocations, a screw dislocation and a circular dislocation loop.
Numerical results show a high consistency in displacement field while no singularity appears in the peridynamics model, the interaction force is in agreement with be the Peach-Koehler formula down to the core region and high accuracy can be reached in 3D with limited computation cost.
The proposed model provides a feasible tool for multiscale modeling of dislocations. Though dislocation is modeled as pre-defined displacement jump, it is straightforward to extend the method to model various fracture conditions.
\end{abstract}

\section{Introduction}
\label{sec:intro}

The physical mechanism of plasticity lies in the collective behaviors of massively distributed dislocations.
In mesoscale, dislocation-induced distortions of the stress and displacement fields are fundamental to the prediction of various nonlinear deformation.
Through decades, dislocation models continuously feed a large amount of  mesoscale physical simulations, e.g., crystal plasticity and dislocation dynamics\citep{Shao2014,Li2014,Borodin2015,Schulz2019,Po2019,Alipour2019}.
Compared with phenomenological constitutive models, direct simulation of solid deformation with dislocations involves the microstructure evolution patterns and thus fills the gap across scales during bottom-up multiscale modeling.

As a kind of lattice defects, dislocations represent irregularly arranged atoms along a line in crystals. Since accurate stress and displacement fields are associated with local lattice structure in modeling dislocations, atomistic simulation tools have shed light on capturing detailed dislocation misfit structure in recent years\citep{Salehinia2014,Askari2015,Tucker2015, Ryu2016, Zhu2017,Kim2019,Tiwari2019a,Mayer2019}, which only depend on lattice parameters but are free from predefined dislocation structure.
However, atomistic simulations including density functional theory and molecular dynamics method meet the bottleneck of computational efficiency in predicting the behaviors of large systems. For the purpose of upscaling, one of the most promising solutions is to bridge atomistic tools with continuum or mesoscale models together concurrently \citep{Gracie2009,Xu2015, Xiong2015}. The other method is to passing defect structure features into upscale models hierarchically\citep{Amodeo2016, Chandra2018, Sun2018b, Krasnikov2019}. Whereas, besides the intrinsic nature of material microstructures, a continuum description of dislocations is necessary for both methods, which should share the physical interpretation of continuum mechanics but also be consistent with the stress and displacement fields of atomistic models.

Generally, dislocations in the continuum scale are constructed by directly incorporating the displacement discontinuities in solids.
 One basic model is to view the displacement jump as a constant equal to the Burgers vector. One such dislocation model is Volterra's "cut and glue" model \citep{Volterra1907}. Notwithstanding the mathematical convenience and tractability of classical continuum mechanics, analytical solutions of displacement and stress fields in the linear elasticity framework are singular. Although several elegant numerical schemes were proposed to avoid singularities \citep{Belytschko2007, Gracie2008,Jamond2016, Huang2019,Liang2019}, the infinite energy and force resulting from singularities are still inconsistent with the atomistic models. Similar to numerical methods, the singularity can also be limited mathematically via introducing an artificial "cut-off" parameter \citep{Gavazza1976}.
Within the framework of classical continuum theory, another category of attempts in removing the singularity is conducting a redistribution of the Burgers vector by energy minimization. In the well known Peierls-Nabarro model \citep{Peierls1940, Nabarro1947}, the displacement field is obtained by minimization of the sum of elastic energy and stacking fault energy, which can be interpreted as the existence of unique dislocation core structure. The dislocation core model is crucial in the Peierls-Nabarro model. According to constrains in the minimization procedure, standard core model \citep{LOTHE1992} and isotropic core model \citep{Cai2006} are developed. The latter provides a non-singular and self-consistent analytical solution, available for state-of-the-art dislocation dynamics simulations \citep{Cui2019,Bertin2018,Niu2019}. The Peierls-Nabarro approaches heavily depends on the core region definition, which is tricky to be investigated in experiments or atomistic simulations.

The inconsistency between the atomistic and continuum dislocation models can be attributed to the scale. In the view of bottom-up scaling, two key features of atomistic scale mechanics are distinct from classical continuum mechanics: discreteness and its related nonlocality. As a lattice defect, dislocation forms in the presence of misfit interactions between atoms, yet it is necessary to highlight that the interaction in the atomistic scale is long-range.
 Therefore, the ignorance of nonlocality in the continuum description of dislocations is doubtful. The application of generalized elasticity theories in dislocations has provided promising results in removing singularities, including the Eringen's nonlocal elasticity theory \citep{Eringen2004}, gradient elasticity \citep{Lazar2011, Po2014a, Lyu2017, Po2018, Wang2016h} and micropolar theories \citep{Clayton2006}.
 For example, in Eringen's nonlocal elasticity theory \citep{Eringen2004} the singularities in the stress field are removed though the singularities in the displacement field remain, suggesting that the singularity is a result of classical continuum theory but not only of the structure of dislocations. Meanwhile, applications of the generalized elasticity theories still suffer from a lack of robust solution techniques even numerically and only recently isogeometric analysis made it hopeful \citep{Rudraraju2014}.

As alluded above, nonlocality is a key to avoid the singularities caused by dislocations.
Amongst numerous generalized continuum theories, peridynamics is a nonlocal theory developed in the last two decades \citep{Silling2000}. In peridynamics, nonlocality is introduced via a reformulation of classical continuum theory. Instead of partial differential equations, the governing equation of peridynamics appears in an integral form to describe internal state variables, which overcomes the singularity problems encountered in discontinuities and thus can be viewed as a coarse grain model upscaled from molecular dynamics \citep{Seleson2009}.
Opposite to other coarse grain methods \citep{Xiong2011,Xiong2014, Deng2010}, peridynamics employs macroscale measurable material parameters so that the tricky choice of miscellaneous atomistic potential functions and other temperature-related properties is avoided. The underlying relationship between peridynamics and atomistic models suggests the potential application in multiscale modeling \citep{Tong2016}.
Up to now, peridynamics has shown great potential in modeling mesoscale defects \citep{Wang2018e} but little work has been done in modeling dislocations. One of the main reasons is the insufficient treatment of discontinuity. In previous studies of discontinuities with peridynamics, which mainly focused on the simulation of crack propagation, the discontinuities were simply assumed as the vanishing of some pairwise interactions passing through. Unfortunately, the assumption has led to different material properties near the discontinuities or surface compared with the bulk part, which is called the surface effect or skin effect \citep{Le2018}. Corrections of the surface effect near boundaries have been widely investigated and greatly improved the accuracy. However, the surface effect near new surfaces or internal discontinuities is still lack of effective control \citep{Le2018}.
In this paper, we introduce an embedded discontinuity method to extend the state-based peridynamics \citep{Silling2010} theory into simulating dislocations with Volterra's dislocation geometry. The state-based peridynamics model is free from the problem of fixed Poisson's ratio in the original bond-based model. The proposed embedded discontinuity method can handle dislocation induced discontinuities without triggering the surface effect and is well-suited in the meshless numerical  framework \citep{Silling2005}. 
To the authors' knowledge, till now this is the only method which can totally remove the surface effect for interior interfaces.
Results indicate that in peridynamics theory both stress and displacement field are regularized. By introducing an interaction range parameter with clear physical interpretation, the peridynamics provides a flexible framework bridging the atomistic models and classical continuum models.

The paper is organized as follows.
In Section \ref{sec:method}, we introduce the representation of the dislocation in the continuum firstly, then the theory of the state-based peridynamics is briefly reviewed, and a constitutive model with embedded discontinuities is derived in Section \ref{sec:edm}. The numerical discretization framework and solution process is the next in Section.\ref{sec:discretization}. The last part is the numerical examples for different types of dislocations, Section \ref{sec:examples}.

\section{Methodology}
\label{sec:method}
In this section, the state-based peridynamics theory is briefly reviewed after defining a continuum description of dislocation, and then we give an explanation that why the discontinuity should be embedded in constitutive rule. Later, details about the construction method of dislocations in peridynamics are described based on the modified Cauchy-Born rule.
\subsection{Definition of solids with  dislocations in continuum}
\begin{figure}[!htp]
	\centering
	\includegraphics[width=0.975\textwidth]{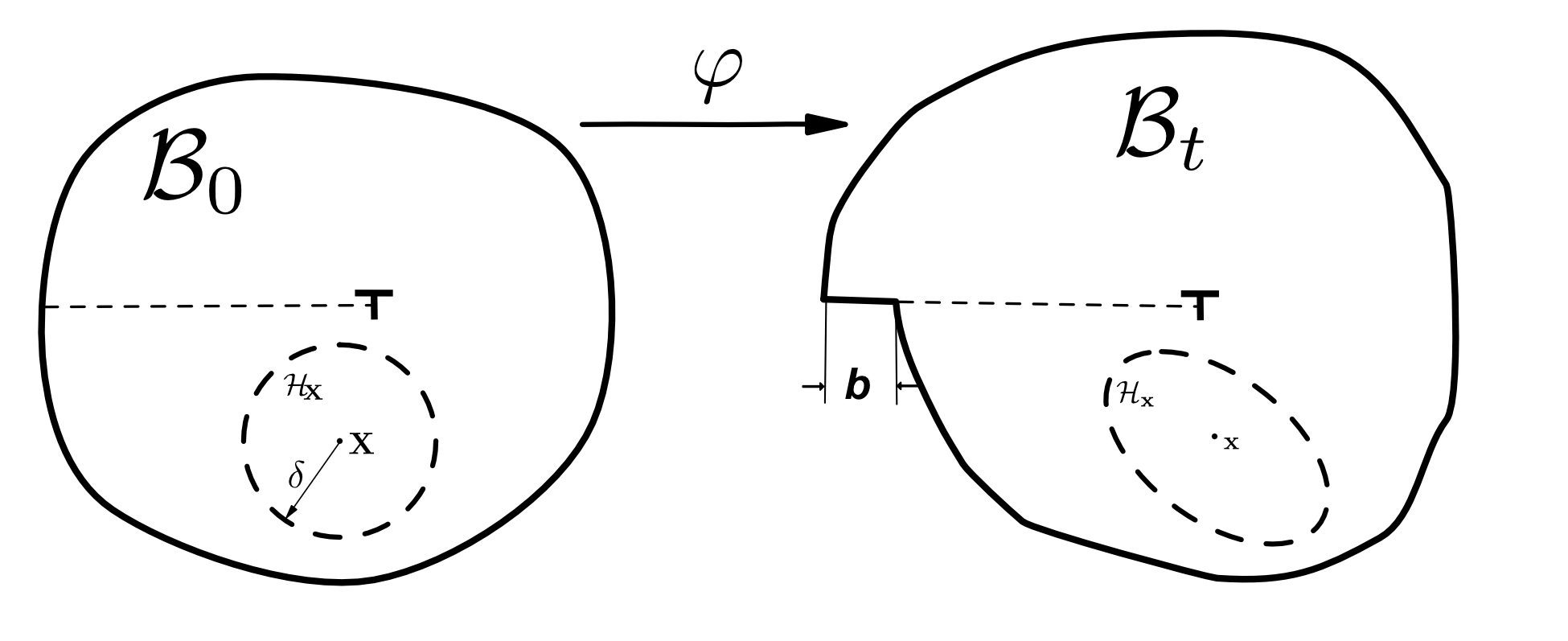}
  	\caption{A solid body contains a Volterra dislocation mapped from reference configuration to deformed configuration. }
  	\label{fig:glide}
\end{figure}
We consider a Volterra dislocation in this work. For brevity, an edge dislocation with the Burgers vector $\mathbf{b}$ is sketched in Fig.\ref{fig:glide}. The dislocation is characterized with the core point position in 2D (dislocation line in 3D) and the glide plane. In the reference configuration $\mathcal{B}_0 \subset \mathbb{R}^n$, $n=2,3$, the glide plane of dislocation is modeled as an interface $\Gamma$ inside the solid body while the core is denoted as $\partial\Gamma$. The interface $\Gamma$ cuts into the solid body and introduces two new surfaces, denoted as $\Gamma^+$ and $\Gamma^-$.
At the time $t$, the material point $\mathbf{X}\subset \mathcal{B}_0$ is mapped to the deformed configuration $\mathcal{B}_t$,
\begin{equation}
	\mathbf{x} = \varphi(\mathbf{X}) \rightarrow \mathbf{u} =  \mathbf{x} - \mathbf{X},
\end{equation}
where $\mathbf{u} $ is the displacement. For a pair of conjugated material points $\mathbf{X}^+\subset \Gamma^+$ and $\mathbf{X}^-\subset \Gamma^-$ defined as $\mathbf{X}^+=\mathbf{X}^-$, the map creates a jump condition across the glide plane,
\begin{equation}
	\mathbf{x}^+ - \mathbf{x}^-= \varphi(\mathbf{X}^+)- \varphi(\mathbf{X}^-)
\quad
\text{or }\quad
	\llbracket{\mathbf{u}}\rrbracket= \mathbf{u}^{+} - \mathbf{u}^{-}.
\end{equation}
For dislocations, the displacement jump is constrained tangent to the glide plane and can be quantified as the Burgers vector $\bm b$. In the Volterra's model, we further assume that $\bm b$ is a constant for the displacement jump across the glide plane of a certain dislocation. Since the introduction of dislocations divided the whole domain of interest into the bulk part $\mathcal{B}$ and the internal interfaces $\Gamma$, the deformation is not homogeneous, which breaks the Cauchy-Born rule and finally cause the state-based peridynamics insufficient for interfaces.
\subsection{The state-based peridynamics}
The state-based peridynamics model is a general theoretical framework of continuum mechanics. According to the assumption of interaction direction constraints, the state-based peridynamics can be split into ordinary \citep{Silling2007} and nonordinary \citep{Warren2009} models. 
In this work, the framework of dislocations is developed based on the linear peridynamic solids model \citep{Silling2010}, which is one of the ordinary models and has distinguished numerical stability compared with the nonordinary models \citep{Ganzenmuller2015a}.
For an arbitrary material point $\mathbf{X}$ in the reference configuration $\mathcal{B}_0$,  the basic assumption of peridynamics is that any point $ \mathbf{X'}$ within a finite distance $\delta$ of $\mathbf{X}$ in $\mathcal{B}_0$ may exert a force upon $\mathbf{X}$.  The interaction distance is denoted as the horizon $\delta$, and the set of interaction points is denoted as the neighbor of $\mathbf{X}$, i.e. $\mathcal{H}_\mathbf{x}$. Thus the balance law is written as
\begin{equation}
	\rho(\mathbf{X}) \mathbf{\ddot u}(\mathbf{X}, t) =
	\int_{\mathcal{H}_\mathbf{X}} \mathbf{f} ( \mathbf{X}, \mathbf{X'},  \mathbf{u}(\mathbf{X}, t), \mathbf{u}(\mathbf{X'}, t)) \mathrm{d} \mathbf{X'}
	+\mathbf{g}(\mathbf{X},t),
\end{equation}
where $\rho$ is the density, $\mathbf{u}$ is the displacement and $\mathbf{g}$ is the body force density. $\mathbf{f} ( \mathbf{X}, \mathbf{X'},  \mathbf{u}(\mathbf{X}, t), \mathbf{u}(\mathbf{X'}, t))$  is the pairwise force density exerted on $\mathbf{X}$ from an point $\mathbf{X'}$ within the horizon $\delta$. In peridynamics, the constitutive modeling is established based on bond stretch measurement. In $\mathcal{B}_0$, an undeformed bond is defined as
\begin{equation}
	\bm{\xi}_{\mathbf{X}\mathbf{X'}} := \mathbf{X'} - \mathbf{X}.
\end{equation}

In bond-based peridynamics, the force density $\mathbf{f}$ between separate points $\mathbf{X'} $ and $\mathbf{X} $ only depends on the behavior of the bond $\bm{\xi}_{\mathbf{X}\mathbf{X'}}$. Unlike the bond-based model, the state-based peridynamics assumes that the force function $\mathbf{f}$ is determined by the collective bonds behavior of the neighbor. Herein, the state is a mathematical object describing the mapping from a collection of variables of the neighbors to a scalar or vector-valued quantity of a specific point, similar to the usage of tensors in classical continuum mechanics. Thus, the concept of the state provides a tool to link the nonlocal model with classical well studied constitutive laws. The pairwise force density exerted on $\mathbf{X}$ in state-based peridynamics is divided into two parts: the force vector state at $\mathbf{X}$ and $\mathbf{X'}$,
\begin{equation}
	\mathbf{f} ( \mathbf{X}, \mathbf{X'},  \mathbf{u}(\mathbf{X}, t), \mathbf{u}(\mathbf{X'}, t)) = \state{\mathbf{T}}{\mathbf{X}, t}{\bm{\xi}_{\mathbf{X}\mathbf{X'}}} - \state{\mathbf{T}}{\mathbf{X'}, t}{\bm{\xi}_{\mathbf{X'}\mathbf{X}}}.
\end{equation}
The underline notation here is denoted as a state. The bracket $[\bullet]$ shows the material point at which it is defined. The angle bracket means that it operates on the the bond $\bm{\xi}$. In ordinary state-based model, it is further assumed that the force vector is collinear with the bond connecting neighbor pairs in $\mathcal{B}_t$. The result is a force density vector pointing to $\mathbf{x}^\prime$ from $\mathbf{x}$ in the deformed configuration $\mathcal{B}_t$. Using the deformed bond vector state, the deformation of bond  ${\bm{\xi_{\mathbf{X}\mathbf{X'}}}}$ can be written as,
\begin{equation}
	\state{\mathbf{Y}}{\mathbf{X},t}{\bm{\xi_{\mathbf{X}\mathbf{X'}}}}
	= \mathbf{x'} - \mathbf{x}.
\end{equation}
Because of the collinear assumption in the ordinary state-based peridynamics, the force vector state can be further decomposed into a scalar-valued force state and a deformed direction vector state,
\begin{equation}
	\state{\mathbf{T}}{\mathbf{x}, t}{\bm{\xi}_{\mathbf{X}\mathbf{X'}}}
	= \state{\rm{T}}{\mathbf{x}, t}{\bm{\xi}_{\mathbf{X}\mathbf{X'}}} \state{\mathbf{M}}{\mathbf{x}, t}{\bm{\xi}_{\mathbf{X}\mathbf{X'}}},
\end{equation}
where $\mathbf{M}$ is the deformed direction vector state, and the value is a unit vector pointing from $\mathbf{x'}$ to $\mathbf{x}$ in $\mathcal{B}_t$,
\begin{equation}
	\state{\mathbf{M}}{\mathbf{X}, t}{\bm{\xi}_{\mathbf{X}\mathbf{X'}}}
	= \dfrac{\mathbf{x'}  - \mathbf{x}}{\|\mathbf{x'}  - \mathbf{x}\|}
	= \dfrac{
	\state{\mathbf{Y}}{\mathbf{X},t}{\bm{\xi_{\mathbf{X}\mathbf{X'}}}} }
	{\|\state{\mathbf{Y}}{\mathbf{X},t}{\bm{\xi_{\mathbf{X}\mathbf{X'}}}} \|}.
\end{equation}
Compared with other upscaling models from molecular dynamics, an important advantage of peridynamics is to incorporate classical continuum constitutive models. The calibration of the scalar force vector in peridynamics utilizes the strain energy density of classical continuum models. Since the modeling of dislocations is in the mesoscale, the material is assumed to be elastic. The deformation of a specific material point is measured by the extension scalar state, defined as
\begin{equation}
	\state{e}{\mathbf{X}, t}{\bm{\xi}_{\mathbf{X}\mathbf{X'}}}  = \| \mathbf{x'}  - \mathbf{x}\| -  \| \mathbf{X'}  - \mathbf{X}\|,
\end{equation}
or using the state notation
\begin{equation}
	\state{e}{\mathbf{X}, t}{\bm{\xi}_{\mathbf{X}\mathbf{X'}}}  = \state{y}{\mathbf{X}, t}{\bm{\xi}_{\mathbf{X}\mathbf{X'}}}  -  \state{x}{\mathbf{X}, t}{\bm{\xi}_{\mathbf{X}\mathbf{X'}}}.
\end{equation}
Here, $\state{y}{\mathbf{X}, t}{\bm{\xi}_{\mathbf{X}\mathbf{X'}}}$ and $\state{x}{\mathbf{X}, t}{\bm{\xi}_{\mathbf{X}\mathbf{X'}}}$ are the magnitude of $\state{\mathbf{Y}}{\mathbf{X}, t}{\bm{\xi}_{\mathbf{X}\mathbf{X'}}}$ and $\state{\mathbf{X}}{\mathbf{X}, t}{\bm{\xi}_{\mathbf{X}\mathbf{X'}}}$ respectively. For brevity, the $[\bullet]$ and $\langle\bullet\rangle$ parts are neglected in the following contents, and it refers to ${[\mathbf{X}, t]}{\langle\bm{\xi}_{\mathbf{X}\mathbf{X'}}\rangle}$ by default.

In peridynamics, the common way for deriving the constitutive relation is via the definition of a strain energy density function $W(\underline{e})$, and the scalar-valued force state is expressed as the Frechet derivative of strain energy density,
\begin{equation}
	\underline{\rm T} = \nabla W(\underline{e}).
\end{equation}
	However, the definition of $W$ in previous literature depends highly on an intact spherical neighbor, which leads to the surface effect when the discontinuities exist in the neighbor. In the next part, we directly find a nonlocal strain energy density function for solid bodies containing interior discontinuities instead of explicit penalty methods, as reviewed by \citet{Le2018}.
\subsection{Constitutive modeling with embedded discontinuity method}\label{sec:edm}
In the previous work \citep{Silling2007,Le2014}, the Cauchy-Born rule is used to build a connection between classical local elasticity and the nonlocal system. Though the Cauchy-Born rule has made a great impact in multiscale modeling, certain shortages do exist.
The drawbacks of the Cauchy-Born rule root in the basic hypothesis of uniform deformation field \citep{Liu2008a}.
In the ordinary state-based peridynamics, the application of the Cauchy-Born rule must be under the constraints of homogeneous deformation in order to reproduce the strain energy density of the corresponding local system. As a kind of inhomogeneous deformation, the occurrence of interior discontinuity shall break the energy conservation. Especially in the previous practice of bond-break modeling of fracture, additional energy dissipation will be brought besides fracture energy, finally leading to an ambiguous\citet{Le2018} crack pattern.
Here, we directly start with the modification of the Cauchy-Born rule accounting for interior discontinuities and later apply the modified Cauchy-Born rule to build a nonlocal strain energy function.
\subsubsection{The modified Cauchy-Born rule}
By assuming a homogeneous small deformation, the standard Cauchy-Born rule for a material point $\mathbf{X}$ with a spherical neighborhood is expressed as,
\begin{equation}
\label{eq.cbr1}
	\mathbf{F}_{\mathbf{X}} \bm{\xi}_{\mathbf{X} \mathbf{X}^\prime}=\mathbf{x}^\prime -\mathbf{x}.
\end{equation}
Here $\mathbf{F}$ denotes the deformation gradient tensor in classical continuum mechanics, $\mathbf{F} = \mathbf{I} + \nabla\mathbf{u}$ and  $\mathbf{I}$ is the identity tensor. It shall be noted that the Cauchy-Born rule requires a smooth enough deformation gradient field in the nonlocal theory. Particularly in the ordinary state-based peridynamics, the nonlocal interaction also requires
\begin{equation}
\label{eq.cbr2}
	\mathbf{F}_{\mathbf{X}^\prime} \bm{\xi}_{\mathbf{X}^\prime \mathbf{X}}=\mathbf{x}-\mathbf{x}^\prime.
\end{equation}
Given that $ \bm{\xi}_{\mathbf{X}^\prime \mathbf{X}} = -  \bm{\xi}_{\mathbf{X} \mathbf{X}^\prime}$, combining Eq.\ref{eq.cbr1} and Eq.\ref{eq.cbr2}, the following must hold,
\begin{equation}
	(\mathbf{F}_{\mathbf{X}} - \mathbf{F}_{\mathbf{X}^\prime}) \bm{\xi}_{\mathbf{X} \mathbf{X}^\prime} = 0
	\quad
	\forall\ \mathbf{X}^\prime \in \mathcal{H}_{\mathbf{X}}.
\end{equation}
Therefore, the above condition would work within acceptable errors only in a small and affine deformation field,
\begin{equation}
\label{eq:homo}
	\mathbf{F}_{\mathbf{X}}  \approx  \mathbf{F}_{\mathbf{X}^\prime}
	\quad
	\text{or}
	\quad
		\mathbf{F}_{\mathbf{X}}  =  \mathbf{F}_{\mathbf{X}^\prime}.
\end{equation}

The ordinary state-based peridynamics is built upon the above assumption.
In other words, the Cauchy-Born rule is based on a small homogeneous deformation field. For inhomogeneous or finite deformation condition, Eq.\ref{eq.cbr1} and Eq.\ref{eq.cbr2} break down.
In the view of displacement discontinuity, the dislocation is a special form of inhomogeneous deformation. In order to model dislocation, we may assume the Cauchy-Born rule is still valid for material points whose neighbor is not cut by the glide plane. For material points near the glide plane, it is also assumed that the Cauchy-Born rule is workable for bonds not intersecting the glide plane. But for bonds intersecting the glide plane, the standard Cauchy-Born rule need modifications to recover the deformation and strain energy defined in classical elasticity.
\begin{figure}[!hbt]
	\centering
	\includegraphics[width=0.975\textwidth]{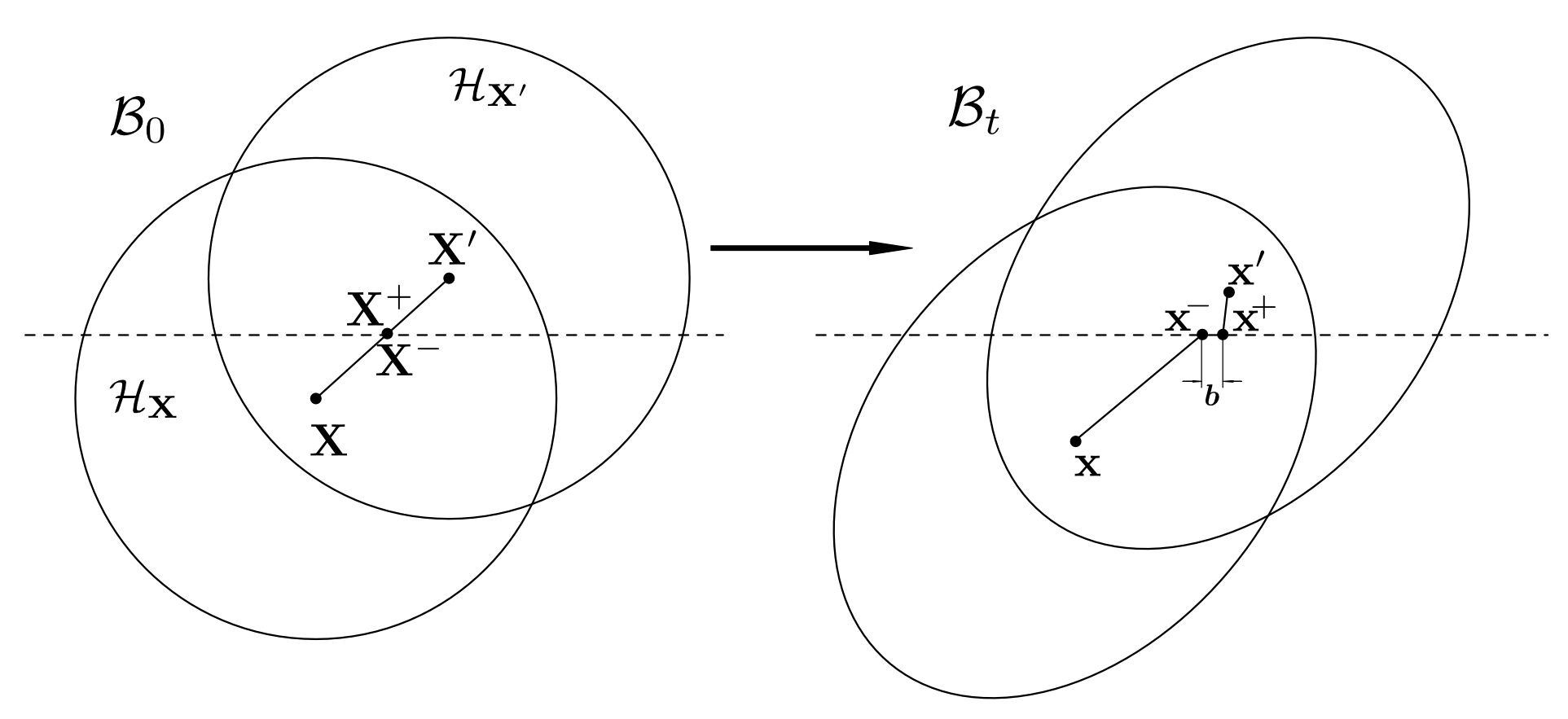}
  	\caption{Incompatibility for interaction bond crossing discontinuity. The dashed line is the glide plane.}
  	\label{fig:discontinuity}
\end{figure}
As shown in Fig.\ref{fig:discontinuity}, consider a pair of adjacent points $\mathbf{X}^{\pm}$ lying on two sides of the glide plane $\Gamma^{\pm}$ in $\mathcal{B}_0$ respectively and its corresponding position $\mathbf{x}^{\pm}$ in $\mathcal{B}_t$, the jump condition induced by dislocations is described with the Burgers vector $\bm{b}$, expressed as,
\begin{equation}
	\bm{b}
	= \mathbf{x^+} -  \mathbf{x^-}.
\end{equation}
Apply the standard Cauchy-Born to bond $\bm{\xi}_{\mathbf{X}\mathbf{X}^-}$ and $\bm{\xi}_{\mathbf{X}^\prime\mathbf{X}^+}$,
\begin{equation}
\label{eq.cbr4}
\left\{
		\begin{aligned}
			&\mathbf{F}_{\mathbf{X}^\prime} \bm{\xi}_{\mathbf{X}^\prime\mathbf{X}^+}
			= \mathbf{x^+} -  \mathbf{x}^\prime \\ \
			&\mathbf{F}_{\mathbf{X}} \bm{\xi}_{\mathbf{X}\mathbf{X}^-}
			= \mathbf{x^-} -  \mathbf{x}
		\end{aligned}
	\right.
\end{equation}
Combine Eq.\ref{eq.cbr4},
\begin{equation}
	\mathbf{F}_{\mathbf{X}} \bm{\xi}_{\mathbf{X}\mathbf{X}^-}  - \mathbf{F}_{\mathbf{X}^\prime} \bm{\xi}_{\mathbf{X}^\prime\mathbf{X}^+}
	=  \mathbf{x}^\prime -\mathbf{x} - \bm{b}.
\end{equation}
Given that $ \bm{\xi}_{\mathbf{X}^\prime \mathbf{X}} = -  \bm{\xi}_{\mathbf{X} \mathbf{X}^\prime}$, and apply the homogeneous deformation condition Eq.\ref{eq:homo}, the modified Cauchy-Born rule with discontinuity can be written as
\begin{equation}
\label{eq:embeddedform}
	\mathbf{F}_{\mathbf{X}} \bm{\xi}_{\mathbf{X}\mathbf{X}^\prime}
	=  \mathbf{x}^\prime -\mathbf{x} - \bm{b}.
\end{equation}
The modification still assumes a small affine deformation field, and by introducing the prescribed displacement jump, the Cauchy-Born rule is still valid. Eq.\ref{eq:embeddedform} can also be interpreted that $\mathbf{x}^+$ and $\mathbf{x}^-$ are replace by $\mathbf{x}$, $\bm{b}$ and $\mathbf{x}^\prime$, relaxing the discontinuity to a nonlocal region. 
\subsubsection{Embedded-discontinuity method}
Similar with the classical mechanics, the nonlocal strain energy density $W$ of the linear peridynamics solid consists the volume part and the distortion part
\begin{equation}
\label{eq.energy}
	W\left(\theta, \underline{e}^{\mathrm{d}}\right)
	=
	\frac{k^{\prime} \theta^{2}}{2}+\frac{\alpha}{2}\left({\omega} \underline{e}^{\mathrm{d}}\right) \bullet \underline{e}^{\mathrm{d}}.
\end{equation}
Here, $k^\prime$ and $\alpha$ are material constants to be calibrated. The notation $\bullet$ is the dot product between two states (cf. \citet{Silling2007}).
$\omega$ denotes the influence function which only depends on the magnitude of $\bm{\xi}$, and in this work we employ the polynomial form proposed by \cite{Seleson2016},
\begin{equation}
\omega(|\bm{\xi}|)
=
\left
	\{\begin{array}{cc}
		1
		-35\left(\frac{|\bm{\xi}|}{\delta}\right)^{4}
		+84\left(\frac{|\bm{\xi}|}{\delta}\right)^{5}
		-70\left(\frac{|\bm{\xi}|}{\delta}\right)^{6}
		+20\left(\frac{|\bm{\xi}|}{\delta}\right)^{7} ,& {|\bm{\xi}| \leqslant \delta},\\
	 	{0} ,& {\text { otherwise }}.
	 \end{array}
\right.
\end{equation}
Corresponding to the classical mechanics, in Eq.\ref{eq.energy} the extension scalar state $\underline{e}$ is divided into two states for the deformation measurement: the nonlocal volume dilatation $\theta$ and the deviatoric extension state $\underline{e}^{\mathrm{d}}$, $\underline{e^d} = \underline{e} - \frac{\theta \underline{x}}{3}$.

\begin{figure}[!hbt]
	\centering
	\includegraphics[width=0.975\textwidth]{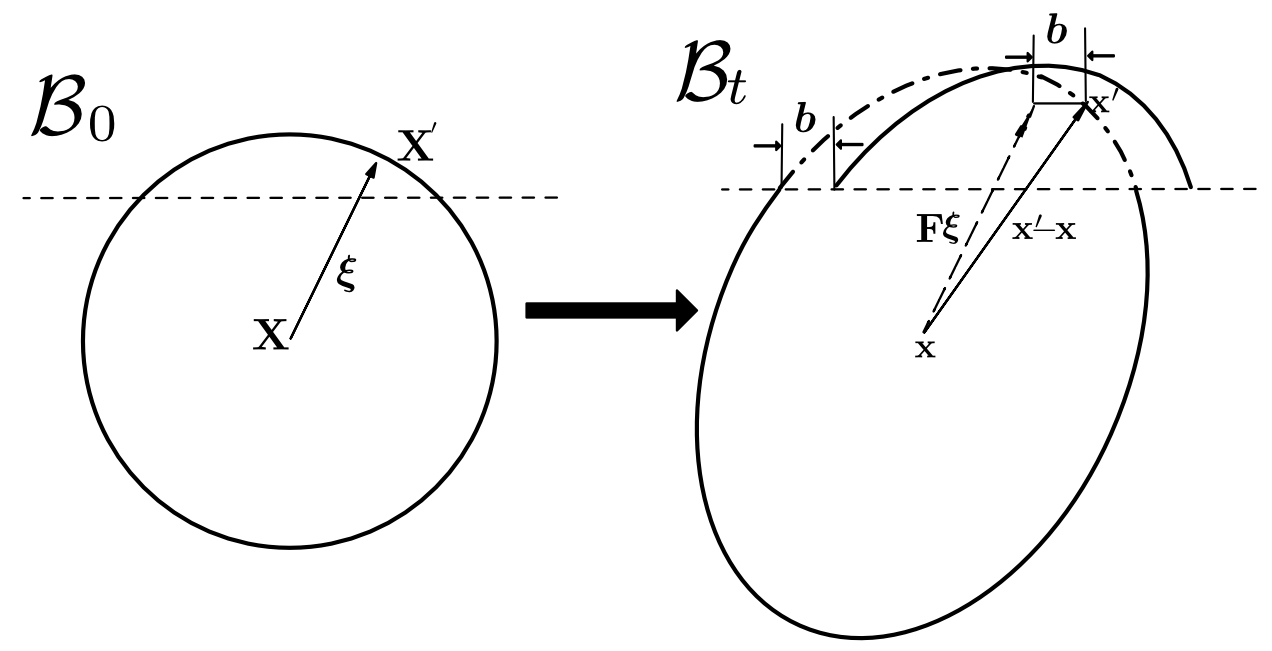}
  	\caption{Illustration of embedded-discontinuity bond. Dashed line: glide plane. The modified Cauchy-Born rule is viewed as the slip of the bond intersecting the glide plane. }
  	\label{fig:deformhorizon}
\end{figure}

The modified Cauchy-Born rule gives an effective way to incorporate a prescribed discontinuity into the modeling process. Similar approach can be found in \citet{Liu2008a, Urata2017}. In the view of nonlocal interaction, the modified Cauchy-Born rule can be understood as interface-induced bond refraction. A bond intersecting the glide plane is shown in Fig.\ref{fig:deformhorizon}. It indicates that the dislocation-induced discontinuity breaks the neighborhood into two sectors, and a relative slip exists in the interface. The modified Cauchy-Born rule can be viewed as shifting the upper sector back to rebuild the continuity of the neighbor. It suggests that the implementation of the modified Cauchy-Born rule in peridynamics is simple, i.e. an embedded-discontinuity deformed bond vector state,
\begin{equation}
	\state{\mathbf{Y^\prime}}{\mathbf{X},t}{\bm{\xi_{\mathbf{X}\mathbf{X'}}}}
	=\state{\mathbf{Y}}{\mathbf{X},t}{\bm{\xi_{\mathbf{X}\mathbf{X'}}}}
	- \sum\bm{b}^\alpha \mathcal{G}^\alpha(\mathbf{X}, \mathbf{X}^\prime)
,
\end{equation}
where $\mathcal{G}^\alpha$ is used for checking the intersection between line segment $\mathbf{X} \mathbf{X}^\prime$ and the glide plane $\Gamma$ of the $\alpha$th dislocation, defined as,
\begin{equation}
	\mathcal{G}^\alpha (\mathbf{X}, \mathbf{X}^\prime)
=
	\left\{
		\begin{aligned}
			1 & \quad  \text{if}\ \mathbf{X} \mathbf{X}^\prime \cap \Gamma^\alpha \neq  \emptyset ,\\
			0 & \quad  \text{otherwise}.
		\end{aligned}
	\right.
\end{equation}
Similarly, the embedded-discontinuity extension scalar state is expressed as
\begin{equation}
	\underline{e}^{\prime} = |\underline{\mathbf{Y}^\prime}| -  |\underline{\mathbf{X}}|.
\end{equation}

The extension scalar state is used for the description of bond elongations. According to the modified Cauchy-Born rule, the length change of a bond $\bm\xi$ is expressed as
\begin{equation}
\label{eq.strain}
	\underline{e}^{\prime} = |\mathbf{F} \bm\xi |-|\bm\xi| = \frac{1}{|\bm{\xi}|}\varepsilon_{ij}\xi_{i}\xi_{j},
\end{equation}
where $\bm{\varepsilon}$  is the infinitesimal strain tensor in classical continuum mechanics. So far, the nonlocal dilatation $\theta$  is given by directly relating to classical volume dilatation. Since the only difference is to replace $\underline{e}$ with $\underline{e}^{\prime}$ between Eq.\ref{eq.strain} and the original work in \citet{Silling2007, Le2014}, the modified nonlocal dilatation $\theta^\prime$ is expressed as
\begin{equation}
\label{eq.dilatation}
	\theta^\prime =
	\left\{
		\begin{aligned}
			\frac{3}{m} (\underline{\omega x}) \bullet \underline{e^\prime
			} & \quad \text{in 3D}, \\
			\frac{2(2\nu-1)}{(\nu-1)m}  (\underline{\omega x}) \bullet \underline{e^\prime} & \quad \text{in plane stress}, \\
			\frac{2}{m}  (\underline{\omega x}) \bullet \underline{e^\prime} & \quad  \text{in plane strain}.
		\end{aligned}
	\right.
\end{equation}
Here, $\nu$ is the Poisson's ratio, and $m$ is the weighted volume defined as $\underline{\omega}\underline{x} \bullet \underline{x}$.
For the deviatoric part of the extension state, a similar equation can be written,
\begin{equation}
\label{eq.ed}
	\underline{e^\prime}^d
	= \underline{e}^{\prime} - \frac{\theta^\prime \underline{x}}{3}.
\end{equation}
Till now, the only difference between the embedded discontinuity model and linear peridynamics model is to replace $\underline{e}$ with $\underline{e}^{\prime}$. Thus deformation energy Eq.\ref{eq.energy} can be expressed with respect to $\underline{e}^{\prime}$ and $\theta^\prime$,
\begin{equation}
\label{eq.energy2}
	W^\prime \left(\theta^\prime, \underline{e^\prime}^{\mathrm{d}}\right)
	=
	\frac{k^{\prime} {\theta^\prime}^{2}}{2}+\frac{\alpha}{2}\left({\omega} \underline{e^\prime}^d \right) \bullet \underline{e^\prime}^d,
\end{equation}
where $k^\prime$ and $\alpha$ are material parameters, as calibrated in \citep{Silling2007, Le2014},
\begin{align*}
	k' =
	\left\{
		\begin{array}{lr}
			k &\text{in 3D}, \\ \vspace{1ex}
			k+\dfrac{\mu}{9} \dfrac{(\nu+1)^2}{(2\nu-1)^2}&\text{in plane stress,}\\
			k + \dfrac{\mu}{9}&\text{in plane strain}.
		\end{array}
	\right.
\\
\\
	\alpha =
	\left\{
		\begin{array}{lr}  \vspace{1.5ex}
			\dfrac{15\mu}{m} &\text{in 3D}, \\
			\dfrac{8\mu}{m} 	&\text{in plane stress or plane strain}.
		\end{array}
	\right.
\end{align*}
Here $k$ is the bulk modulus. By Frechet derivate of $W$ with respect to $ \underline{e}^{\prime}$, the modified scalar force state $\underline{\mathrm T}^\prime $ is written as
\begin{equation}
\label{eq.scalar}
	\Delta W = k^\prime \theta^\prime
	( \nabla_{\underline{e}^{\prime}} \theta^\prime)  \bullet \Delta \underline{e}^{\prime}
	+
	\alpha ({\omega}\underline{e^\prime}^d)
	\bullet
	\Delta \underline{e^\prime}^d
	=\underline{\mathrm T^{\prime}} \bullet \Delta \underline{e}^{\prime}.
\end{equation}
Substitue Eq.\ref{eq.ed} and Eq.\ref{eq.dilatation} into Eq.\ref{eq.scalar},
\begin{equation}
	\underline{\mathrm T^{\prime}} =
	\left\{
		\begin{array}{ll} \vspace{1.5ex}
			3k'\theta^\prime \dfrac{\underline{\omega x}}{m} +
			\alpha {\underline{\omega {e^{\prime}}^d}}& \text{in 3D}, \\ \vspace{1.5ex}
			\dfrac{2(2\nu-1)}{\nu -1}(k'\theta^\prime - \dfrac{\alpha}{3}({\underline{\omega {e^{\prime}}^d}})\bullet \underline{x})\dfrac{\underline{\omega x}}{m} + \alpha {\underline{\omega {e^{\prime}}^d}} & \text{in plane stress},  \\
			2(k'\theta^\prime - \dfrac{\alpha}{3}({\underline{\omega{e^{\prime}}^d}})\bullet \underline{x})\dfrac{\underline{\omega x}}{m} + \alpha {\underline{\omega {e^{\prime}}^d}}  & \text{in plane strain}.
		\end{array}
	\right.
\end{equation}
The force vector state is modified as
\begin{equation}
	\state{\mathbf{T}}{\mathbf{x}, t}{\bm{\xi}_{\mathbf{X}\mathbf{X'}}}
	= \state{\rm{T^{\prime}}}{\mathbf{x}, t}{\bm{\xi}_{\mathbf{X}\mathbf{X'}}} \state{\mathbf{M^{\prime}}}{\mathbf{x}, t}{\bm{\xi}_{\mathbf{X}\mathbf{X'}}},
\end{equation}
and the modified deformed direction vector state is,
\begin{equation}
	\state{\mathbf{M}^\prime}{\mathbf{X}, t}{\bm{\xi}_{\mathbf{X}\mathbf{X'}}}
	= \dfrac{\state{\mathbf{Y^\prime}}{\mathbf{X},t}{\bm{\xi_{\mathbf{X}\mathbf{X'}}}}}
	{\|\state{\mathbf{Y^\prime}}{\mathbf{X},t}{\bm{\xi_{\mathbf{X}\mathbf{X'}}}} \|}.
\end{equation}
\begin{rmk}
In the classical practice of fracture modeling with peridynamics, the crack type strong discontinuity is represented as the "break" of the bond. The failure criterion is expressed with the stretch ratio of the bond, simple and effective. However, after bond-breaking the standard Cauchy-Born rule is unable to reproduce the strain energy of the remaining part, where the neighbor is not spherical. The surface effect \citep{Le2018} appears in the previous bond-based and ordinary state-based peridynamics is the result of such undesired energy loss. Instead, the embedded discontinuity peridynamics directly modifies the interaction force for bonds crossing the discontinuity. Thus the embedded discontinuity method is a conceptually different approach. The modification in energy can be explained as a superposition of the strain energy density created by the perfectly smooth displacement field and the dissipation energy induced by the dislocation interface. In the sense of force, the model can also be understood as the mechanism that additional force states distributed along the glide plane are added as body force to force the displacement jump at the magnitude of the Burgers vector.
\end{rmk}
\subsection{Numerical discretization}
\label{sec:discretization}
\begin{figure}[!hbt]
	\centering
	\includegraphics[width=0.73\textwidth]{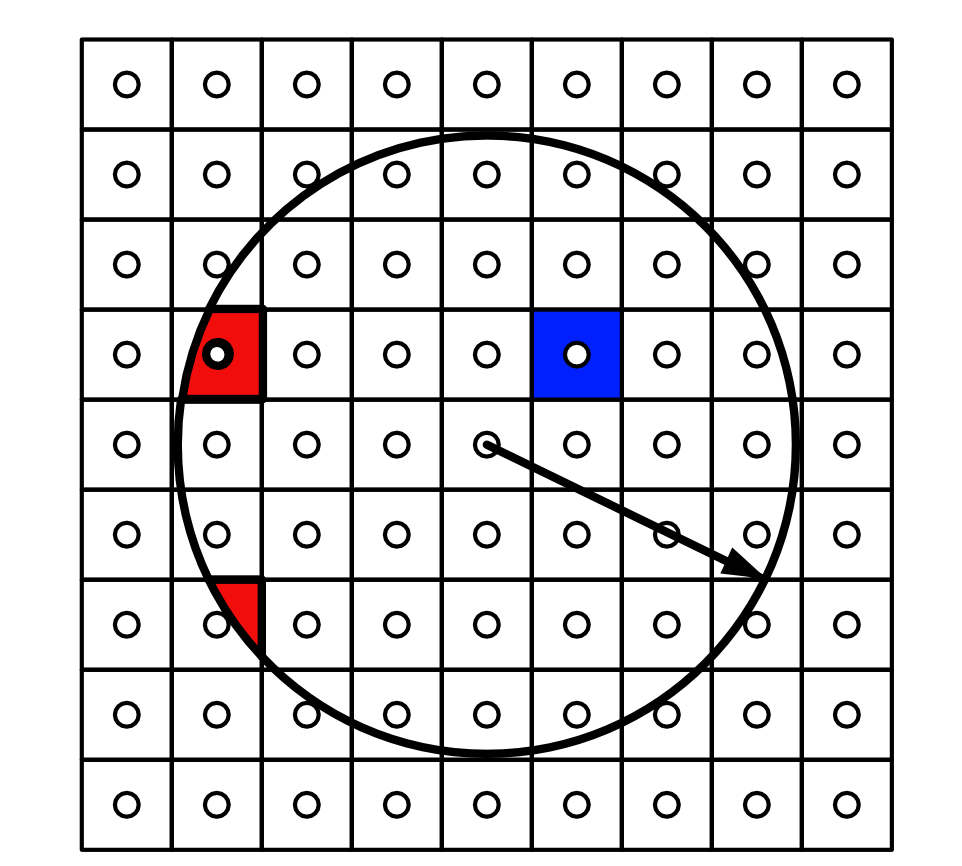}
  	\caption{Illustration of discretization. Red: partial volume. Blue: full volume.}
  	\label{fig:volume}
\end{figure}

Due to the nonlinear nature of peridynamics, analytical solution can seldom be found. Hence in this work we simulate dislocations by the particle-based meshless numerical approach \citep{Silling2005}. The domain is discretized with equal-spaced nodes, as shown in Fig.\ref{fig:volume}. Thus the total force acting on a node can be calculated with Riemann sum, written as,
\begin{equation}
\label{eq.discrete}
\mathbf{F}(\mathbf{X}, t) =
	\sum_{\mathcal{H}_\mathbf{X}} \mathbf{f} ( \mathbf{X}, \mathbf{X'},  \mathbf{u}(\mathbf{X}, t), \mathbf{u}(\mathbf{X'}, t)) \mathrm{V}_ \mathbf{X'}
	+\mathbf{g}(\mathbf{X},t).
\end{equation}
Here $\mathrm{V}_ \mathbf{X'}$ is the volume of each node. For the partially covered volume shown in Fig.\ref{fig:volume}, the IPA-HHB algorithm \citep{Bobaru2012, Seleson2014} is employed to improve the accuracy. 
The boundary conditions in peridynamics has the same physical meaning as in classical elasticity.
The traction boundary condition is applied by reproducing the flux through boundaries and the displacement boundary condition is applied by setting a constant value. The difference is that the boundary condition is applied to a layer but not a lower dimensional geometry.
In this work, the boundary conditions are applied to a layer of nodes whose width equals to the horizon $\delta$. The fictitious boundary layer method \citep{Bobaru2017,Le2018} is used to determine the displacement value for the fictitious nodes. 
The velocity Verlet time integration and the fast inertial relaxation engine method \citep{Bitzek2006} are used for solving static solution. Details of the implement is shown in Algorithm.\ref{alg:fire}. In this paper, we select the parameters for FIRE algorithm as $n_{min}=5$, $\gamma_0=0.1$, $f_{\gamma}=0.99$, $f_{dec}=0.5$, $f_{inc}=1.1$ and the time step is determined with the Courant-Friedrichs-Lewy condition \citep{Silling2005}.

The stress definition in peridynamics is considerably ambiguous in previous literature. Although an elegant definition of the peridynamic stress tensor is given in \citet{LEHOUCQ2008}, widely misuse and meaningless comparisons with the Cauchy stress do exist since the peridynamic stress is corresponding to the Piola stress. Here, we use the mechanical part of the virial stress formula as the equivalent Cauchy stress measurement in the embedded discontinuity peridynamics model, expressed as
\begin{equation}
\label{eq.virial}
	{\bm\sigma}(\mathbf{X} )
	= \frac{1}{2} \int_{\mathcal{H}_\mathbf{X}}
	 \mathbf{f} ( \mathbf{X}, \mathbf{X'},  \mathbf{u}(\mathbf{X}, t), \mathbf{u}(\mathbf{X'}, t))
	 \otimes
	  \state{\mathbf{Y^\prime}}{\mathbf{X},t}{\bm{\xi_{\mathbf{X}\mathbf{X'}}}}
	  \mathrm{d}\mathbf{x'}.
\end{equation}
\begin{algorithm}[H]
  \caption{Static solver }
  \label{alg:fire}
  \begin{algorithmic}[1]
    \Require
      $\mathbf{X}$: node position;
      $\Delta t$: time step size;
      $n_{min}$, $\gamma_0$, $f_{\gamma}$, $f_{dec}$, $f_{inc}$: control parameters;
      $k$, $\nu$: material parameters;
    \Ensure
      optimal $\mathbf{u}$
    \State initial $\gamma = \gamma_0$ and $\Delta t$;
  \While{ not converged}
    \State apply the boundary condition
      \State update the position $\mathbf{x}(t+\Delta t) = \mathbf{x}(t) +  \mathbf{v}(t)\Delta t + \frac{1}{2}\mathbf{a}\Delta t ^2$;
     \State compute the internal force density $\mathbf{F}(t+\Delta t) = \sum_{\mathcal{H}_\mathbf{X}} \mathbf{f} (t+\Delta t) \mathrm{V}_ \mathbf{X'}$;
     \State update the acceleration $\mathbf{a} (t+\Delta t) = \mathbf{F} (t+\Delta t) /\rho$

      \State update the velocity $\mathbf{v} (t+\Delta t) = \mathbf{v}(t) + \frac{1}{2}(\mathbf{a}(t) + \mathbf{a}(t+\Delta t))\Delta t$;
      \State compute the  $P =\mathbf{F}\cdot\mathbf{v}$;
      \State adjust the velocity by 
      $\mathbf{F_N} = \mathbf{F} / \|\mathbf{F_N}\|$;
      $\mathbf{v} \rightarrow (1- \gamma)\mathbf{v} + \gamma \mathbf{F_N} |\mathbf{v}|$;
      \If {$P>0$ and $n>n_{min}$}
      	\State
      	$\Delta t \rightarrow \min({\Delta t f_{inc}, \Delta t_{max}})$;
      	$\gamma \rightarrow\gamma f_{\gamma}$;
      	$n=n+1$;
      \Else
      	\State
      	$\mathbf{v}\rightarrow 0$;
      	$\Delta t \rightarrow \Delta t f_{dec}$;
      	$\gamma \rightarrow\gamma_{0}$;
      	$n = 0$;
      \EndIf
     \EndWhile
      \end{algorithmic}
\end{algorithm}
\section{Numerical examples}
\label{sec:examples}
\subsection{Edge dislocation in an infinite domain}
\label{sec:case1}
\begin{figure}[hbt]
	\centering
	\includegraphics[width=0.5\textwidth]{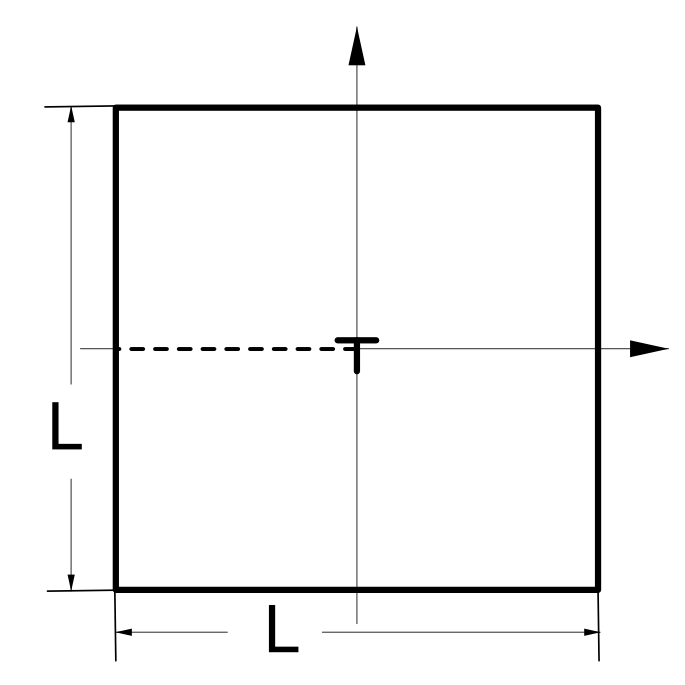}
  	\caption{Illustration of the simulation domain for an edge dislocation}
  	\label{fig:case1}
\end{figure}
A single edge dislocation is considered as benchmark to validate the method. For an infinite domain, the analytical solutions with the Burgers vector $\bm{b}=[b,0,0]$ in the classical elasticity are given in \citet{Hirth1983},
\begin{equation}
\label{eq.u}
	\begin{split}
	 u_{x} & =\frac{b}{2 \pi}\left[\arctan \frac{y}{x}+\frac{x y}{2(1-\nu)\left(x^{2}+y^{2}\right)}\right] ,\\
	 u_{y} & =-\frac{b}{2 \pi}\left[\frac{1-2 \nu}{4(1-\nu)} \ln \left(x^{2}+y^{2}\right)+\frac{x^{2}-y^{2}}{4(1-\nu)\left(x^{2}+y^{2}\right)}\right].
	\end{split}
\end{equation}
The geometry of the simulation domain is a two dimensional square with $L= 10^{-6}\si{\meter}$, and the core of an edge dislocation with $b=8.551\times 10^{-10} \si{m}$ is placed at the origin, as shown in Fig.\ref{fig:case1}.
The Young's modulus and the Poisson's ratio are $1.2141\times10^{11} \si{Pa}$ and $0.34$, respectively. The plane strain condition is assumed. For mimicking an infinite domain, the displacement solution Eq.\ref{eq.u} in classical elasticity is applied to the fictitious boundary layers.
To analyze the effectiveness of meshless discretization, $N \times N$ particles are equally distributed in the domain, and the convergence can be checked with character number $M= {\delta N}/{L}$.
\begin{figure}[!hbt]
	\centering
	\subfloat[][]{
		\label{fig:ux_pd}
			\includegraphics[width=0.487\textwidth]{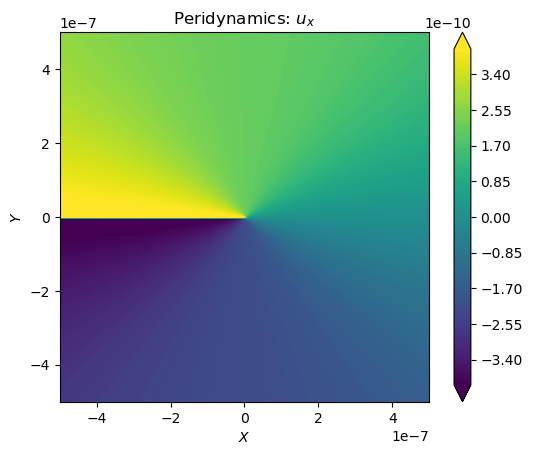}}
	\subfloat[][]{
		\label{fig:ux_a}
			\includegraphics[width=0.487\textwidth]{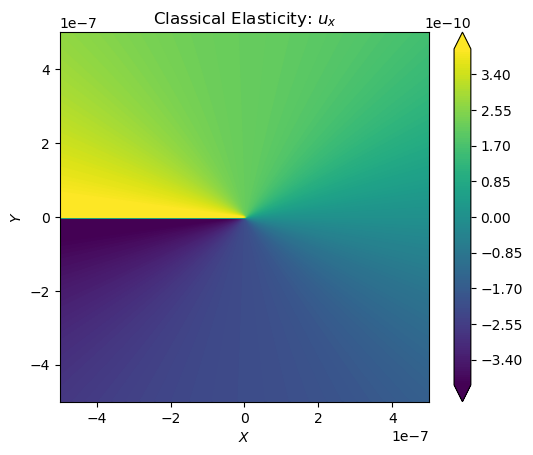}}\\
	\subfloat[][]{
		\label{fig:pd_uy}
			\includegraphics[width=0.487\textwidth]{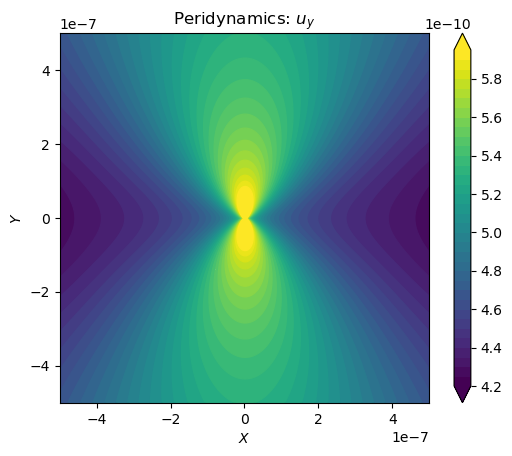}}
	\subfloat[][]{
		\label{fig:a_uy}
			\includegraphics[width=0.487\textwidth]{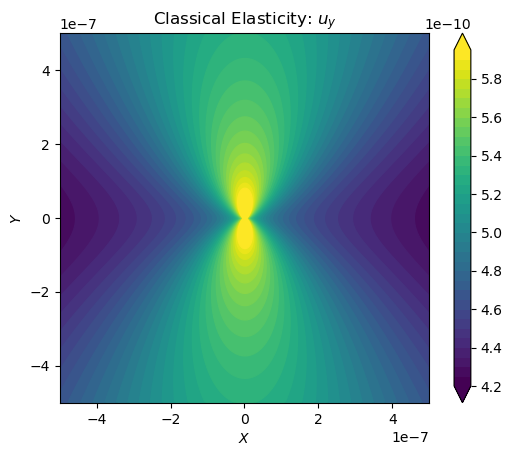}}
\caption[]
{Displacement field induced by an edge dislocation with embedded discontinuity peridynamics \subref{fig:ux_pd}\subref{fig:pd_uy} and classical elasticity\subref{fig:ux_a}\subref{fig:a_uy}.
}
\label{fig:displace_edge}%
\end{figure}

\begin{figure}
	\centering
	\subfloat[][]{
		\label{fig:stressxx_pd}
			\includegraphics[width=0.479\textwidth]{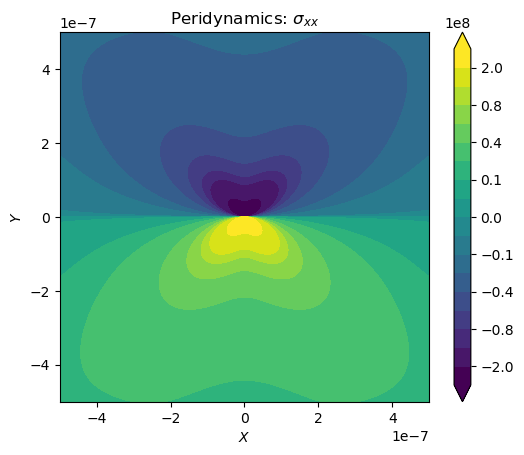}}
	\subfloat[][]{
		\label{fig:stressxx_a}
			\includegraphics[width=0.479\textwidth]{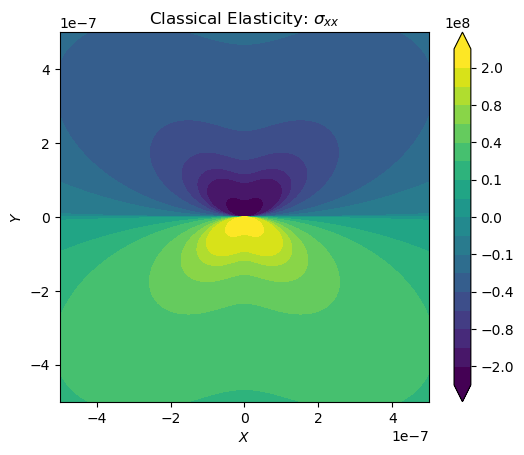}}
			\\
	\subfloat[][]{
		\label{fig:stressyy_pd}
			\includegraphics[width=0.479\textwidth]{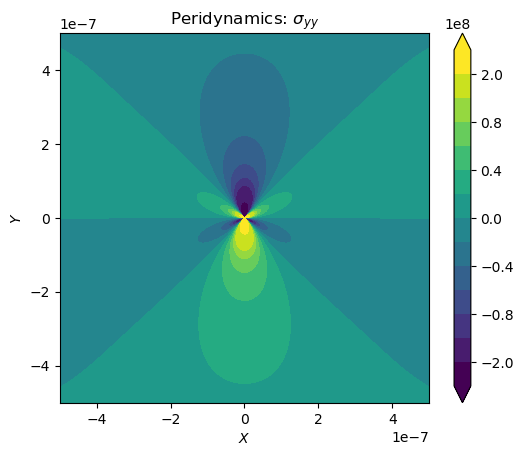}}
	\subfloat[][]{
		\label{fig:stressyy_a}
			\includegraphics[width=0.479\textwidth]{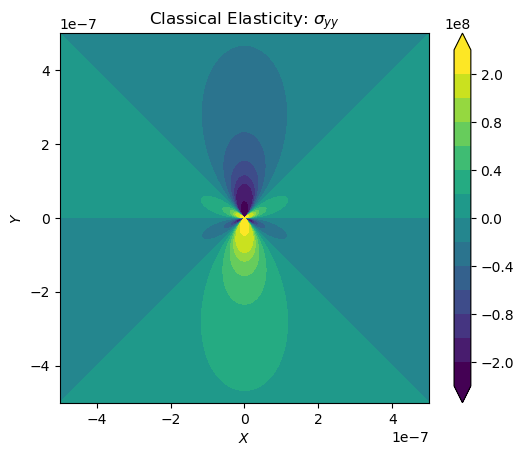}}
\\	
	\subfloat[][]{
		\label{fig:stressxy_pd}
			\includegraphics[width=0.479\textwidth]{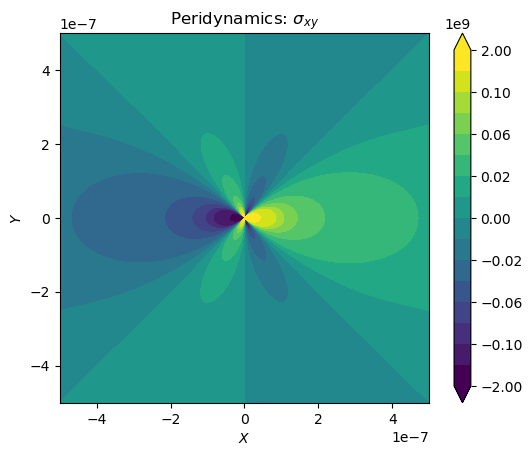}}
	\subfloat[][]{
		\label{fig:stressxy_a}
			\includegraphics[width=0.479\textwidth]{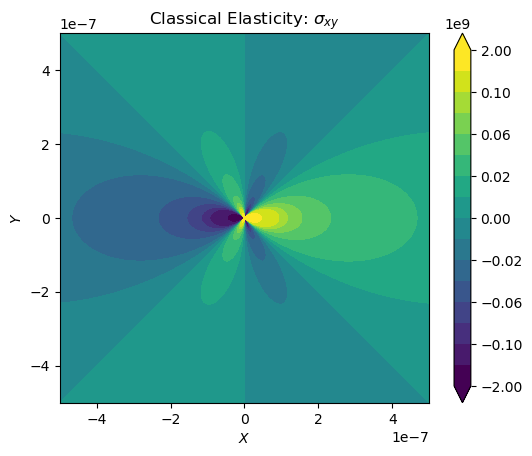}}
\caption[]
{
Stress field induced by an edge dislocation with embedded discontinuity peridynamics \subref{fig:stressxx_pd}\subref{fig:stressyy_pd}\subref{fig:stressxy_pd} and classical elasticity\subref{fig:stressxx_a}\subref{fig:stressyy_a}\subref{fig:stressxy_a}
}
\label{fig:stress_edge}%
\end{figure}

In Fig.\ref{fig:displace_edge} and \ref{fig:stress_edge}, the stress and displacement obtained by setting $N=500$ and $M=3.15$ are compared with the classical elasticity solution.
Obviously, the surface effect common in previous fracture studies \citep{Le2018} disappears. In the proposed embedded discontinuity method, the bulk strain energy is totally reproduced since all bulk points have a full horizon.  The displacement results of peridynamics match well with the classical elasticity, and only a slight difference exists between the stress components. Besides negligible numerical issues, it proves that the nonlocal interaction, or the horizon $\delta$ really redistributes the stress field. The differences of stress in Fig.\ref{fig:stress_edge} are mainly due to three reasons: nonlocality, stress definition and numerical errors. Nonlocality in peridynamics introduces a difference in the solution of displacement field while $\delta>0$. The difference between peridynamics and classical elasticity converges to zero when $\delta$ approaches zero, which is called $\delta$ convergence. Compared with the displacement results in Fig.\ref{fig:displace_edge}, it is obvious that the discrepancy mainly exists in stress results. In this work the mechanical part of virial stress is used as an measurement of the Cauchy stress, which is equivalent but still influenced by the introduction of $\delta$. 
\begin{figure}[!hbt]
	\centering
	\includegraphics[width=1.0\textwidth]{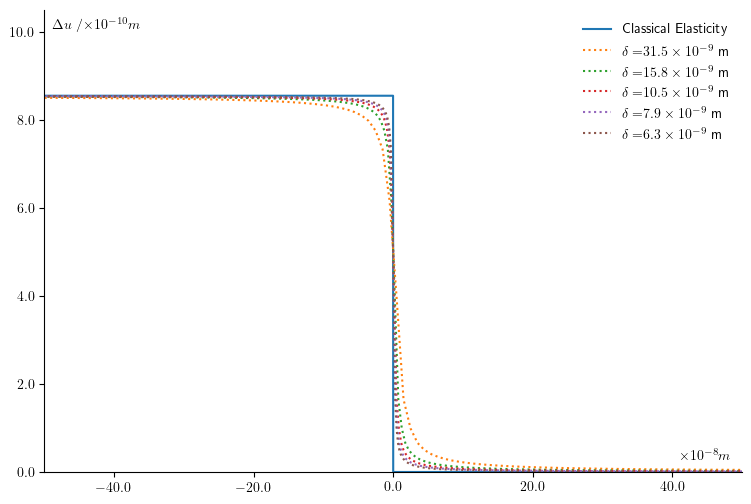}
  	\caption{Displacement jump along the glide plane. Values on the horizontal axis refer to the X coordinate in Fig.\ref{fig:displace_edge}.}
  	\label{fig:jump}
\end{figure}

In Fig.\ref{fig:jump}, the effectiveness of the embedded discontinuity peridynamics in capturing the discontinuity jump is validated. As $\delta$ decreases, the displacement jump curve approaches the classical solution, which confirms the $\delta$-convergence of peridynamics. Besides perfectly recovering the Burgers vector for most regions, in the near-core region the displacement jump or recovered Burgers vector is gradually decreasing. It also appeared in the dislocation model with the gradient elasticity theory \citep{Po2014a}. The phenomenon indicates a redistribution of the Burgers vector with embedded discontinuity peridynamics in the near core region, conceptually similar to the result with the Peierls-Nabarro model and related non-singular theory \citep{Cai2006}.

\begin{figure}[!hbt]
	\centering
	\includegraphics[width=1.0\textwidth]{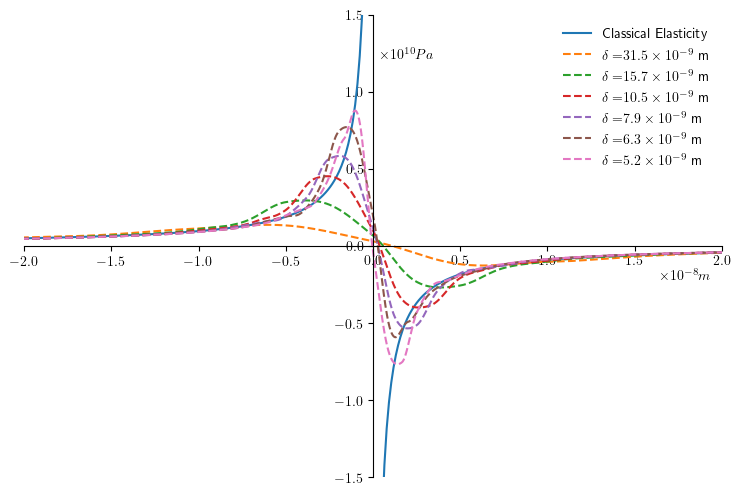}
  	\caption{$\sigma_{xx}$ induced by an edge dislocation. Plotting along the line segment from $(0, -2\times10^{-8} m)$ to $(0, 2\times10^{-8} m)$ is shown. Values on the horizontal axis refer to the X coordinates in Fig.\ref{fig:stress_edge}.}
  	\label{fig:compare}
\end{figure}

\begin{figure}[!hbt]
	\centering
			\includegraphics[width=0.85\textwidth]{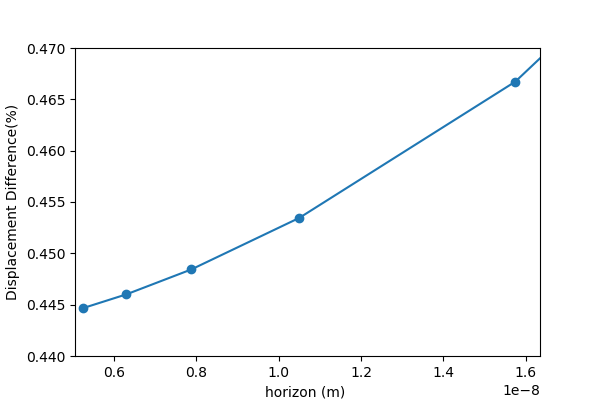}
\caption{Relative displacement difference between embedded discontinuity peridynamics and classical elasticity, calculated with Eq.\ref{eq:udiff}.}
\label{fig:converge}
\end{figure}

Fig.\ref{fig:compare} shows the influence of horizon on the distribution of the stress component $\sigma_{xx}$.
A series of $\delta$ was implemented in numerical simulations with a fixed $N=500$ but varying $M$.
The stress $\sigma_{xx}$ is only plotted for the near core region along the line segment from $(0, -2\times10^{-8} m)$ to $(0, 2\times10^{-8} m)$.
Compared with classical elasticity, the singularity is avoided with the embedded discontinuity peridynamics. The stress curves obtained with different $\delta$ are finite but diverges around the core position. As $\delta \rightarrow0$, the stress curve near the core position gradually rises towards the classical elasticity solution, which is usually described as $\delta$-convergence.
 As an important feature of peridynamics, we characterize the $\delta$-converge rate with the relative $L_2$ norm between the displacement $\mathbf{u}^{\text{num}}$ obtained with the embedded discontinuity peridynamics and the analytical solution $\mathbf{u}^{\text{local}}$ in Eq.\ref{eq.u}, expressed as
\begin{equation}
\label{eq:udiff}
	D_u =
	 \frac{\| \mathbf{u}_{h}-\mathbf{u}_{}\|_2}
	{\|\mathbf{u}_{}\|_2}  
= 
  \sqrt{\frac{\int_{\mathcal{B}}(\mathbf{u}^{\text{num}} - \mathbf{u}^{\text {local}})^{\mathsf{T}}(\mathbf{u}^{\text{num}}-\mathbf{u}^{\text {local}}) 
  				d\mathcal{B}}{\int_{\mathcal{B}} (\mathbf{u}^{\text{local}})^{\mathsf{T}} (\mathbf{u}^{\text{local}}) d \mathcal{B}}} .
\end{equation}
Fig.\ref{fig:converge} shows the $\delta$-convergence of displacement field. With the embedded discontinuity model, the difference of the displacement field between classical elasticity and peridynamics is relatively small, but a rapid decrease is still shown as $\delta \rightarrow 0$. In the embedded discontinuity model, the reproduction of the bulk part of strain energy is guaranteed. Besides avoiding the surface effect, the $\delta$-convergence rate is kept constant even with the presence of discontinuity.

\subsection{Interaction between edge dislocations}
\label{sec:case2}
In this section, the interaction between two edge dislocations is considered. To compare with Section \ref{sec:case1}, we choose the same set of material and dislocation parameters. In geometry, one edge dislocation is still placed at the origin while the other is at $(L_x, L_y)$. In discretization, we choose $N=500$ and $M=3.15$.

\begin{figure}[!hbt]
	\centering
	\subfloat[][$u_x$ at $L_x=0$ \si{m}]{
		\label{fig:double_ux_0}
			\includegraphics[width=0.45\textwidth]{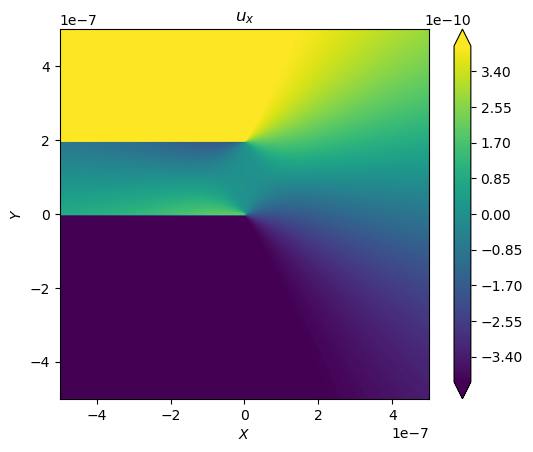}}
	\hspace{1pt}
	\subfloat[][$u_y$ at $L_x=0$ \si{m}]{
		\label{fig:double_uy_0}
			\includegraphics[width=0.45\textwidth]{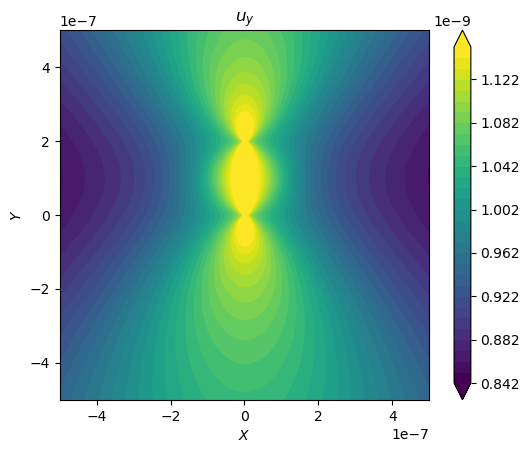}}
\\
	\subfloat[][$u_x$ at $L_x=1\times 10^{-7}  \si{m}$]{
		\label{fig:double_ux_50}
			\includegraphics[width=0.45\textwidth]{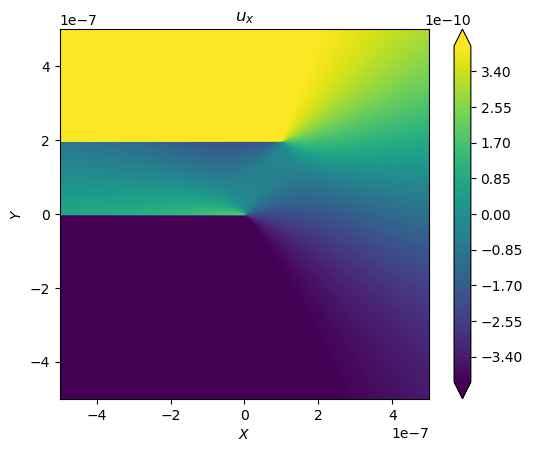}}
	\hspace{1pt}
	\subfloat[][$u_y$ at $L_x=1\times 10^{-7}  \si{m}$]{
		\label{fig:double_uy_50}
			\includegraphics[width=0.45\textwidth]{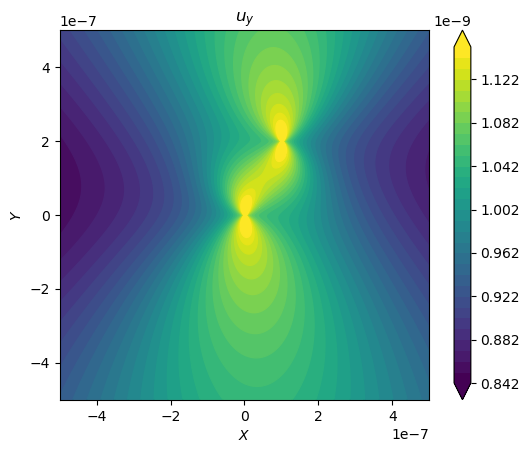}}
\\	
	\subfloat[][$u_x$ at $L_x=2\times 10^{-7}  \si{m}$]{
		\label{fig:double_ux_99}
			\includegraphics[width=0.45\textwidth]{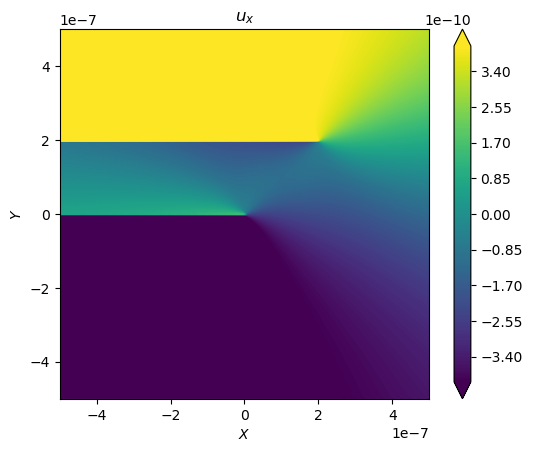}}
	\subfloat[][$u_x$ at $L_x=2\times 10^{-7}  \si{m}$]{
		\label{fig:double_uy_99}
			\includegraphics[width=0.45\textwidth]{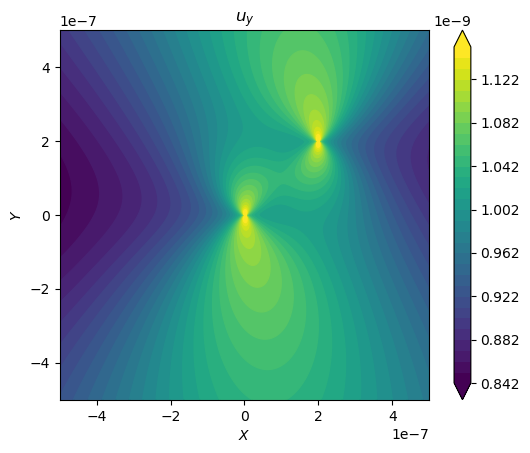}}
\caption[A set of four sub-floats.]
{Displacement field induced by two edge dislocations, unit: \si{m}.
}
\label{fig:double_u}%
\end{figure}

\begin{figure}[!hbt]
	\centering
	\subfloat[][ $\sigma_{xx}$ at $L_x=0 \si{m}$]{
		\label{fig:double_stressxx_0}
			\includegraphics[width=0.3\textwidth]{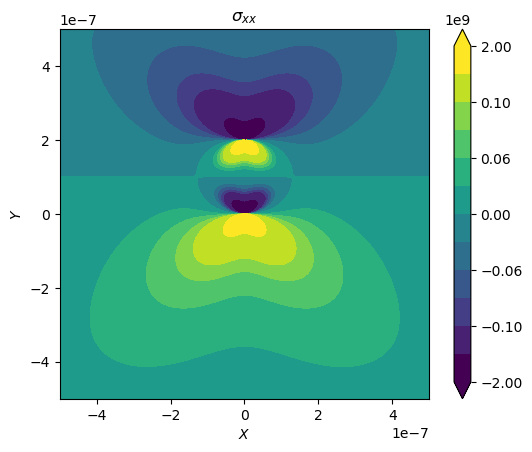}}
	\hspace{1pt}
	\subfloat[][ $\sigma_{xx}$ at $L_x=1\times 10^{-7}  \si{m}$]{
		\label{fig:double_stressxx_50}
			\includegraphics[width=0.3\textwidth]{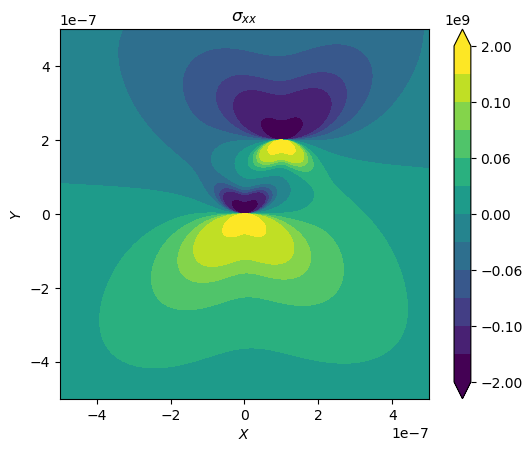}}
	\hspace{1pt}
	\subfloat[][ $\sigma_{xx}$ at $L_x=2\times 10^{-7}  \si{m}$]{
		\label{fig:double_stressxx_99}
			\includegraphics[width=0.3\textwidth]{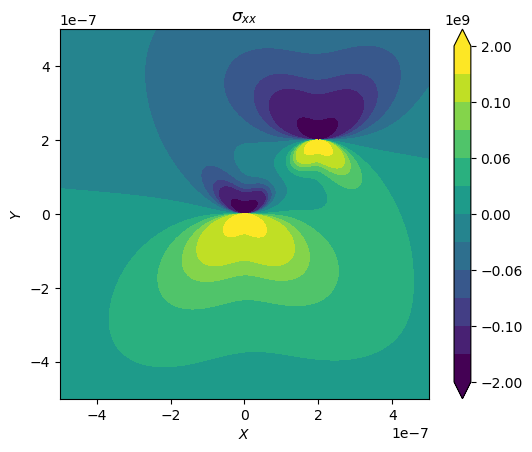}}
	\\
	\subfloat[][$\sigma_{xy}$ at $L_x=0 \si{m}$]{
		\label{fig:double_stressxy_0}
			\includegraphics[width=0.3\textwidth]{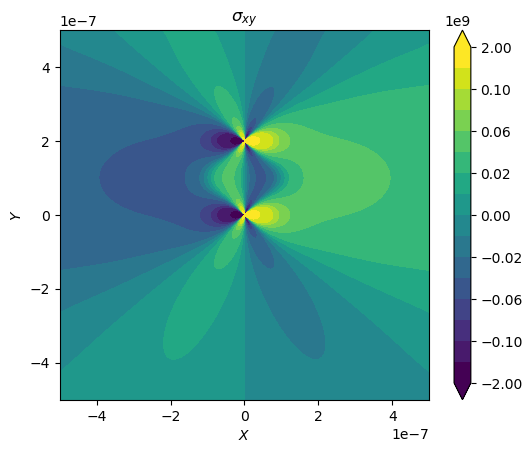}}
	\hspace{1pt}
	\subfloat[][$\sigma_{xy}$ at $L_x=1\times 10^{-7}  \si{m}$]{
		\label{fig:double_stressxy_50}
			\includegraphics[width=0.3\textwidth]{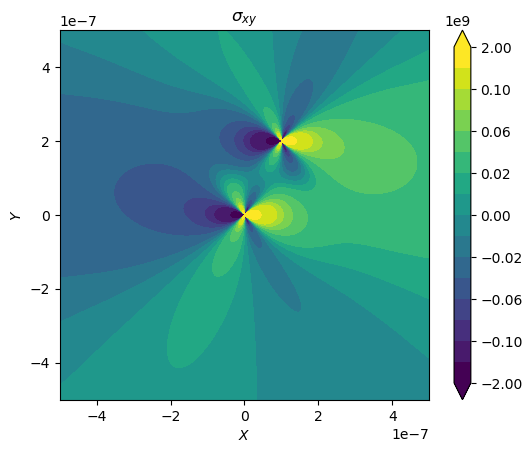}}
	\hspace{1pt}
	\subfloat[][$\sigma_{xy}$ at $L_x=2\times 10^{-7}  \si{m}$]{
		\label{fig:double_stressxy_99}
			\includegraphics[width=0.3\textwidth]{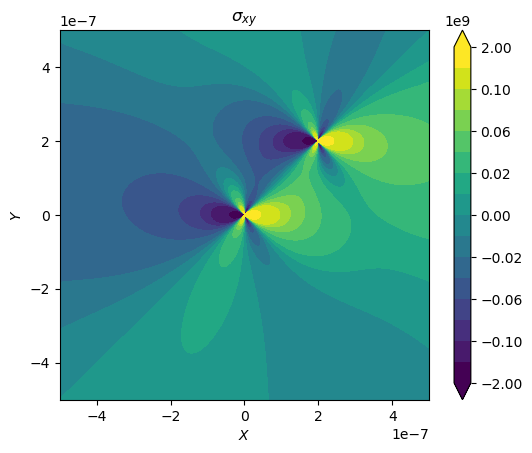}}
	\\
	\subfloat[][$\sigma_{yy}$ at $L_x=0 \si{m}$]{
		\label{fig:double_stressyy_0}
			\includegraphics[width=0.3\textwidth]{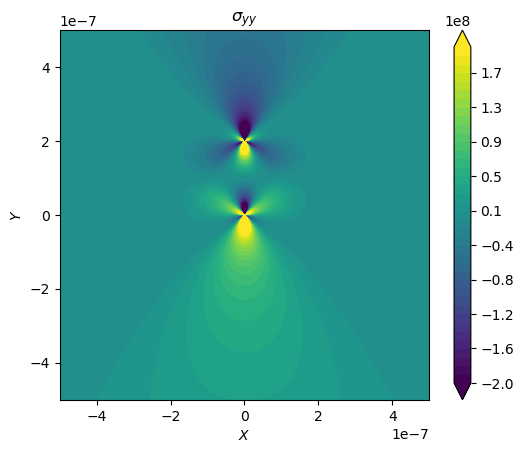}}
	\hspace{1pt}
	\subfloat[][$\sigma_{yy}$ at $L_x=1\times 10^{-7}  \si{m}$]{
		\label{fig:double_stressyy_50}
			\includegraphics[width=0.3\textwidth]{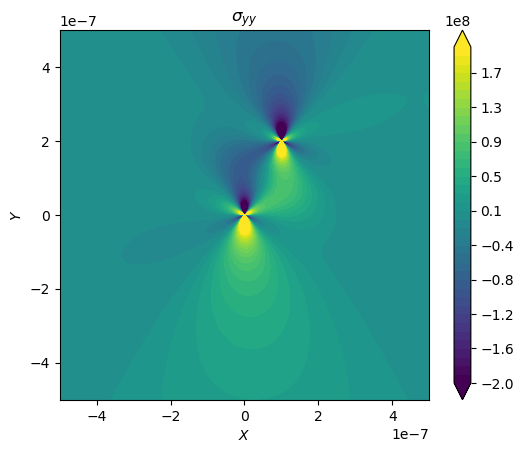}}
	\hspace{1pt}
	\subfloat[][$\sigma_{yy}$ at $L_x=2\times 10^{-7}  \si{m}$]{
		\label{fig:double_stressyy_99}
			\includegraphics[width=0.3\textwidth]{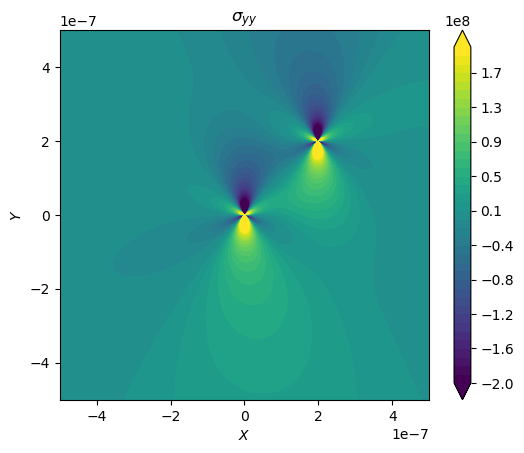}}
	\\
\caption[A set of four sub-floats.]
{Stress field induced by two edge dislocations, color bar unit: \si{Pa}.
}
\label{fig:double_stress}%
\end{figure}
In Fig.\ref{fig:double_u} and \ref{fig:double_stress}, the displacement and the stress fields at the same $L_y=2\times 10^{-7} \si{m}$ but three different $L_x$ are shown. As seen from the figures, both the stress and the displacement field agree well with the classical elasticity result, and  a further quantitive measurement of $D_u$ shows $D_u = 0.27\% \sim 0.33\%$ for  all $L_x = 0 \sim 2\times 10^{-7}  \si{m}$. The results suggest that the application of embedded discontinuity peridynamics in multiple dislocations is feasible. Another issue on the multiple dislocations modeling is the interaction force, or driving force in dislocation dynamics.
In classical elasticity, the driving force $\mathbf{F}$ on unit dislocation line segment is defined as the negative derivatives of elastic strain energy $E$ with respect to the coordinates $\mathbf{x}$,
\begin{equation}
	\label{eq:eg}
	\mathbf{F} = -\dfrac{\partial E}{\partial \mathbf{x}}.
\end{equation}
On the other side, the driving force is also consistent with the Peach-Koehler formula,
\begin{equation}
	\label{eq:pk}
	\mathbf{F} = ( \bm{\sigma} \cdot \bm{b} ) \times \bm{\xi},
\end{equation}
where $\bm{\xi}$ is a unit vector tangent to the dislocation line, and $\bm{\sigma}$ is the Cauchy stress.

\begin{figure}[!hbt]
	\centering
	\includegraphics[width=1.0\textwidth]{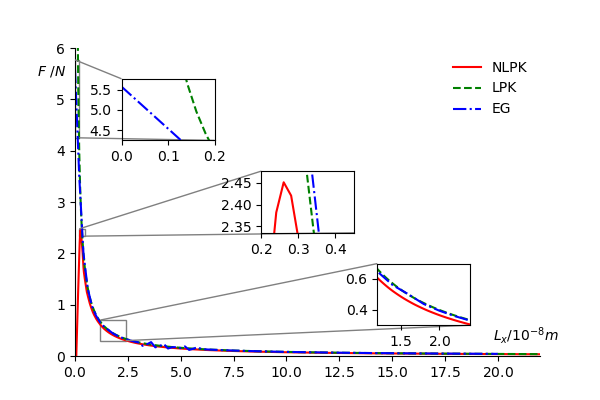}
  	\caption{Driving force on dislocation line.}
  	\label{fig:pkcompare}
\end{figure}

Here, we compute the driving force numerically by three methods,
\begin{itemize}
  \item[\bf{NLPK}] The nonlocal Peach-Koehler force is calculated via substituting the nonlocal stress defined in Eq.\ref{eq.virial} to Eq.\ref{eq:pk},  and the nonlocal stress is interpolated at the corresponding position utilizing the numerical data from Sec.\ref{sec:case1}.
  \item[\bf{LPK}] The local Peach-Koehler force is calculated with the Peach-Koehler formula and the analytical solution in classical elasticity.
  \item[\bf{EG}] The energy gradient method is calculated with Eq.\ref{eq:eg} using second order accurate central differences respect to $L_x$ numerically. A series of simulations was performed by setting  $L_y = 0 \si{m}$ and $L_x= 2n\times 10^{-9} \si{m}$, $n= 0,1\cdots100$.
The energy is computed by directly summing the strain energy density defined in Eq.\ref{eq.energy} for the whole simulation domain.
\end{itemize}
 Fig.\ref{fig:pkcompare} compares the driven force calculated with the above three methods.
With negligible numerical errors, the three methods show a high degree of consistency. Apart from the drawbacks of low order particle meshless method, the numerical errors can also be ascribed to the finite simulation domain compared with the infinite domain solution in the classical elasticity.
The main differences mainly exist in the near core region, where in classical elasticity solution the dislocation core model is not included and the LPK is approaching infinite as $L_x \rightarrow0$. The embedded peridynamics solution avoids the singularity in energy, thus the interaction force is also finite. The NLPK is inconsistent with the EG as $L_x \rightarrow0$. Consider the $\delta = 31.5 \times 10^{-9}$ m, the inconsistency indicates the behavior of dislocation core is described in the embedded discontinuity peridynamics model, and the near core region interaction is failed to be described with the Peach-Koehler formula. It should also be noted that the interaction between edge dislocations within the dislocation core distance is not fully understood in literature so far. Thus the potential use of the embedded discontinuity peridynamics is not limited to dislocation dynamics with the Peach-Koehler formula but can also be extended to study the core region behavior.
\subsection{Screw dislocation in an infinite domain}
\label{sec:case2}
\begin{figure}[!hbt]
	\centering
	\includegraphics[width=0.6\textwidth]{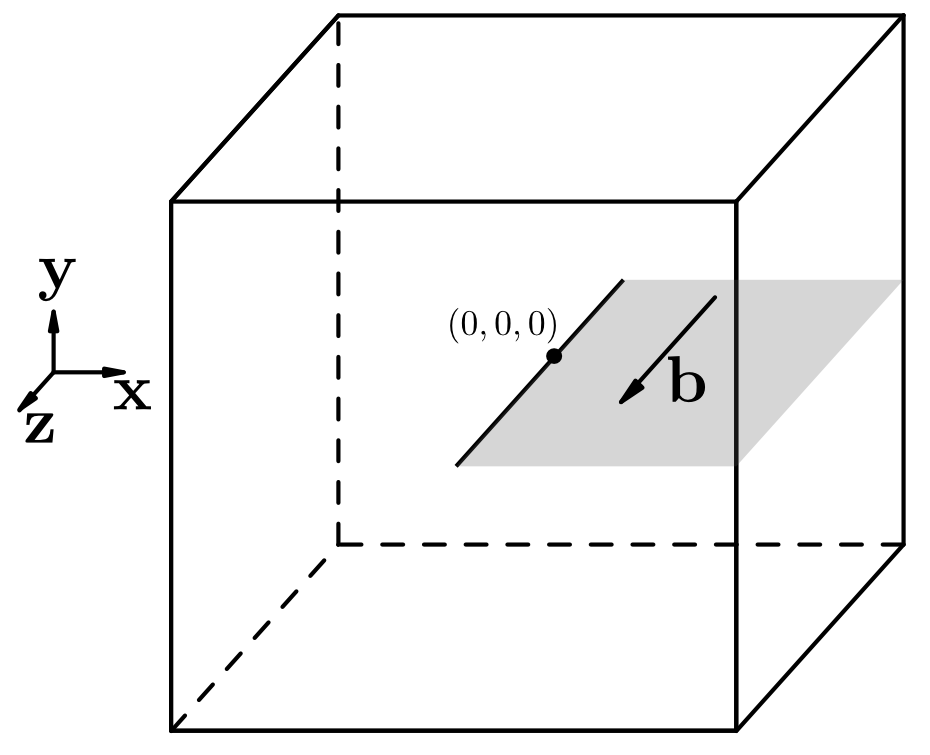}
  	\caption{Illustration of the simulation domain for a screw dislocation}
  	\label{fig:screw}
\end{figure}
To validate the 3D condition, a straight screw dislocation in an infinite domain is considered. The domain geometry is shown in Fig.\ref{fig:screw}. The domain is a cube with edge length $L= 10^{-6}\si{\meter}$, and dislocation line coincides with the $Y$ axis. The Burgers vector $\bm{b}=[0,0,b]$ is selected as $b=8.551\times 10^{-10} \si{m}$. The Young's modulus and the Poisson's ratio are $1.35\times10^{10} \si{Pa}$ and $0.28$, respectively. We set a fixed $M=3.15$ but different $N$ in discretization.
\begin{figure}[!hbt]
	\centering
	\subfloat[][$u_z$]{
		\label{fig:screwuz}
			\includegraphics[width=0.5\textwidth]{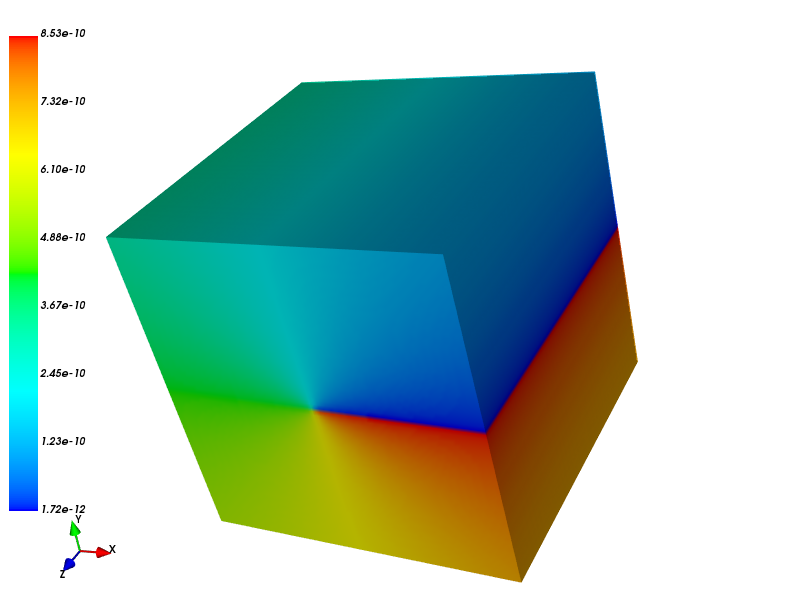}}
			\\
	\subfloat[][$\sigma_{xz}$]{
		\label{fig:screwstressxz}
			\includegraphics[width=0.5\textwidth]{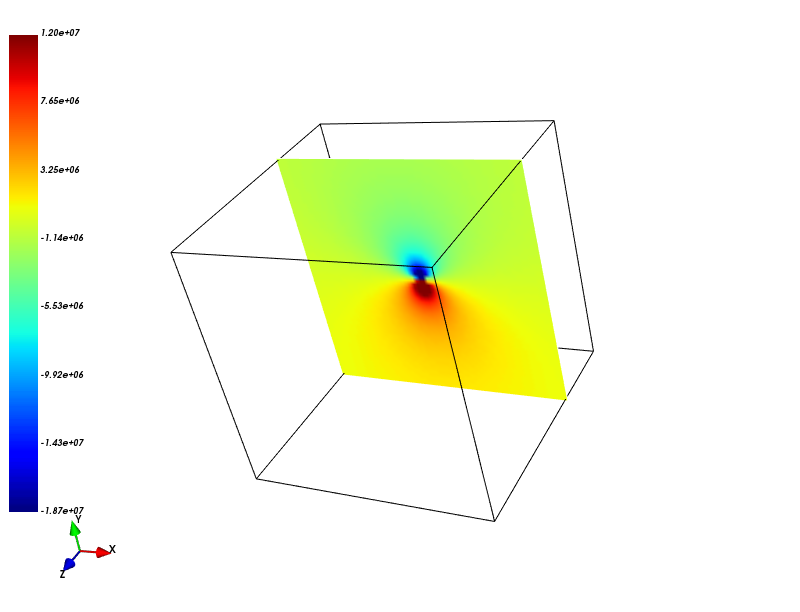}}
	\subfloat[][$\sigma_{yz}$]{
		\label{fig:screwstressyz}
			\includegraphics[width=0.5\textwidth]{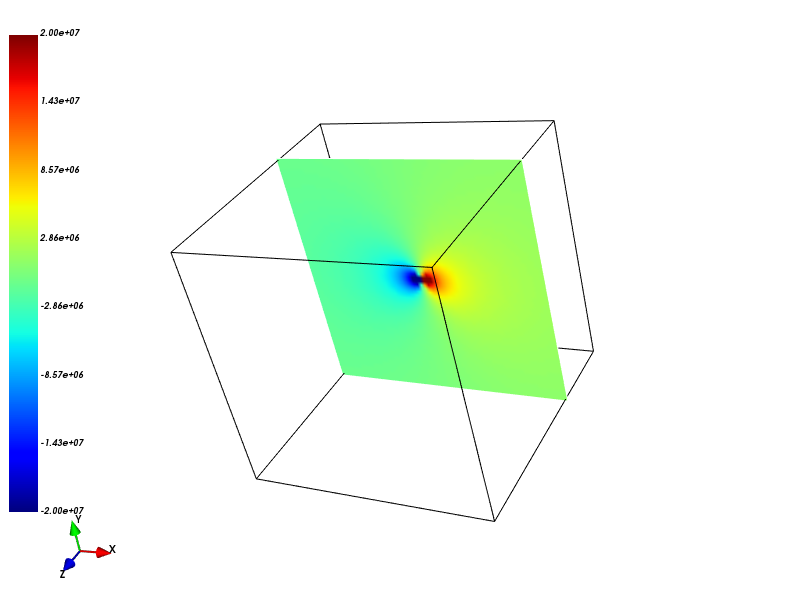}}
\caption[]
{Displacement and stress field induce by a screw dislocation,
\subref{fig:screwstressxz} and
\subref{fig:screwstressyz} describes the stress field of a slice plane at $z = -2.5\times 10^{-7} \si{m}$
}
\label{fig:screwresult}
\end{figure}

\begin{figure}[htp]
	\centering
	\subfloat[][$\sigma_{xz}$]{
		\label{fig:screw_compare_stress}
			\includegraphics[width=0.9\textwidth]{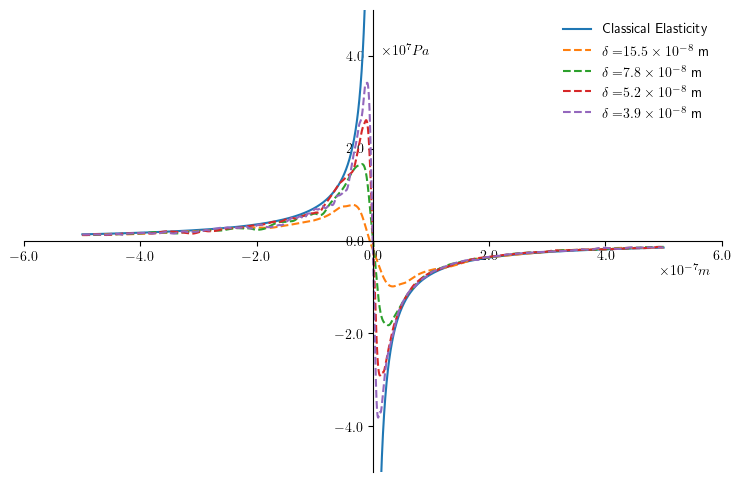}}\\
	\subfloat[][$u_z$]{
		\label{fig:screw_compare_disp}
			\includegraphics[width=0.9\textwidth]{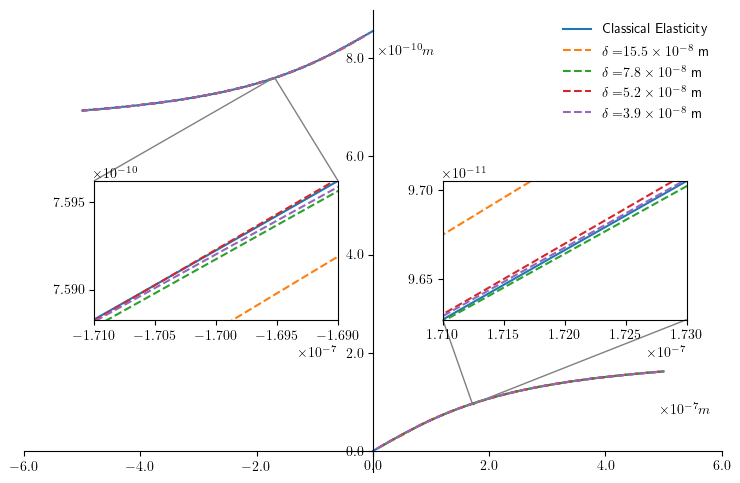}}
\caption[]
{Displacement and stress induced by a screw dislocation.
\subref{fig:screw_compare_stress} stress, plotted from $(0, -5, 0)\times10^{-7}\si{m}$ to $(0, 5, 0)\times10^{-7}\si{m}$ and
\subref{fig:screw_compare_disp} displacement, plotted from $(2, -5, 0)\times10^{-7}\si{m}$ to $(2, 5, 0)\times10^{-7}\si{m}$. Values on the horizontal axis of both subfigures refer to the Y coordinate in Fig.\ref{fig:screw}.
}
\label{fig:screwcompare}
\end{figure}

The displacement and stress field caused by the screw dislocation with $N=80$ is presented in Fig.\ref{fig:screwresult}. Only nonzero components are shown here. All results are independent of the z coordinate by examining an arbitrary slice normal to the Z-axis and are in good accordance with the classical elasticity solution \citep{Hirth1983}.
An in-depth comparison is preformed via plotting the displacement and stress components along the selected line segment with different $\delta$, as shown in Fig.\ref{fig:screwcompare}.
Compared with the classical elasticity solution, the singularity in $\sigma_{xz}$ vanishes for all $\delta$ with embedded discontinuity peridynamics, Fig.\ref{fig:screw_compare_stress}. Similar to the stress of edge dislocation in Fig.\ref{fig:compare}, the stress curve passes through the origin and also gradually converges to the classical elasticity solution as $\delta\rightarrow0$. Thus the $\delta$ convergence is confirmed. The phenomenon is also observed in the nonsingular theory by \citet{Cai2006} and gradient elasticity \citep{Lazar2011}. In all the above models, the regularization of singularity is indeed by the redistribution of local energy into a nonlocal range. The Peierls-Nabarro type model constrains the redistribution to the glide plane or jump condition while the nonlocal models including the peridynamics extend it to the whole domain.
The embedded discontinuity peridynamics can also be proved to be effective by examining the displacement field in Fig.\ref{fig:screw_compare_disp}. Although the stress field is regularized, a high degree of consistency is still maintained for the displacement field. The $\delta$ convergence can also be seen in local enlarged subfigures inside Fig.\ref{fig:screw_compare_disp}. Quantitively, the relative displacement difference keeps decreasing as $\delta\rightarrow0$, shown in Table \ref{table:screw}. Remarkably, the accuracy is achieved with very rough discretization $N$ for reducing computation cost. Thus the embedded discontinuity peridynamics model is also less sensitive in discretization.

\begin{table}
\centering
\caption{Relative displacement difference with different horizon}
\label{table:screw}
\begin{tabular}{lllr}
	\toprule
	Horizon ($\times10^{-8}\si{m}$)& $M$&$N$ & $D_{u}$(\%)\\
	\midrule
		15.5 &3.15 &20 & 0.5327\\
		7.8 & 3.15&40 & 0.2857\\
		5.2 & 3.15&60 &0.1980\\
		3.9 & 3.15&80 &0.1525\\
 	\bottomrule
	\end{tabular}
\end{table}
\subsection{Circular dislocation loop}
	\label{sec:case3}
\begin{figure}[htbp]
		\centering
		\includegraphics[width=0.8\textwidth]{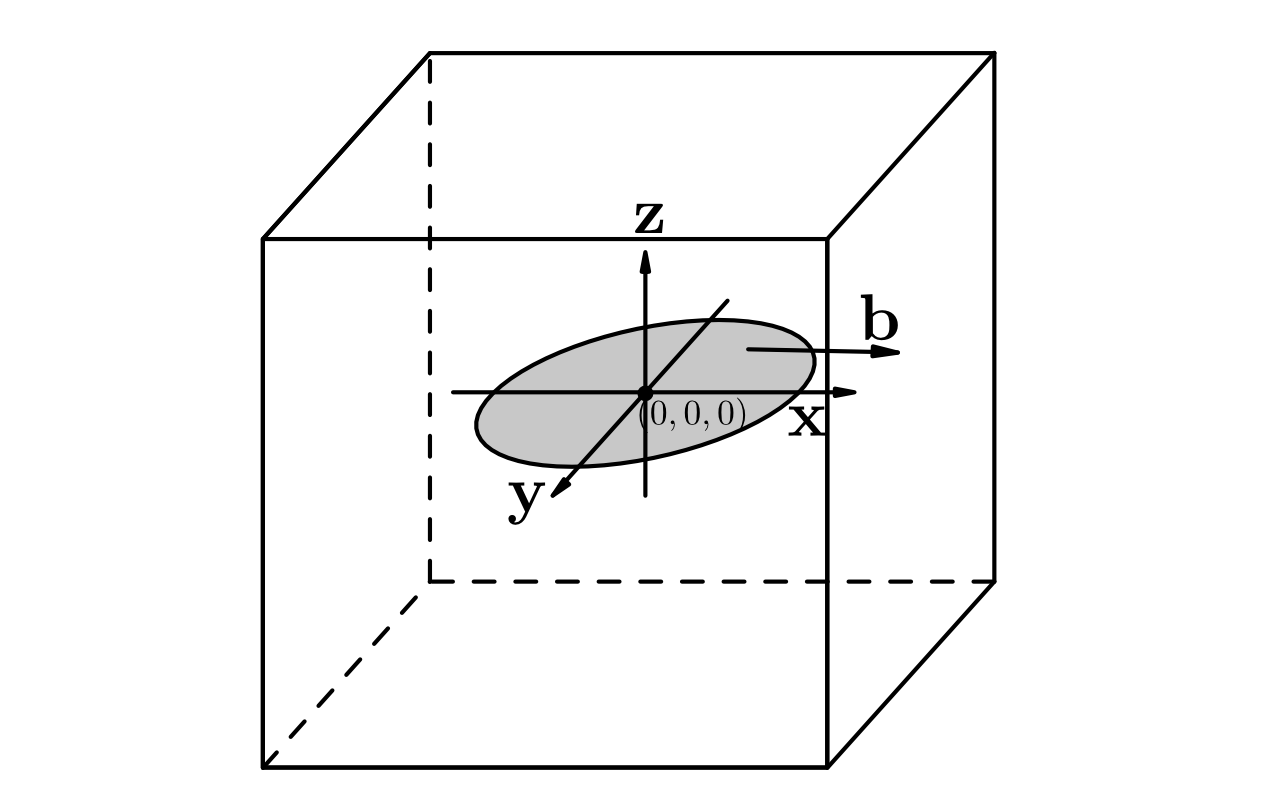}
  	\caption{Illustration of the simulation domain for a circular dislocation loop}
  	\label{fig:loop}
\end{figure}

\begin{figure}[htbp]
	\centering
		\subfloat[][$u_{z}$]{
		\label{fig:loop_compare_dispz}
			\includegraphics[width=0.98\textwidth]{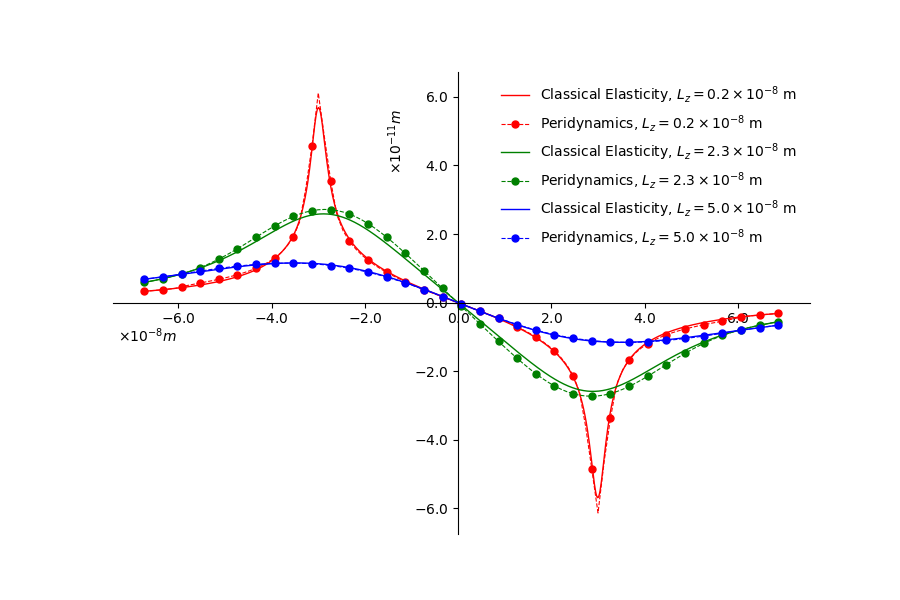}
			}\\
		\subfloat[][$\sigma_{xz}$]{
		\label{fig:loop_compare_stressxz}
			\includegraphics[width=0.98\textwidth]{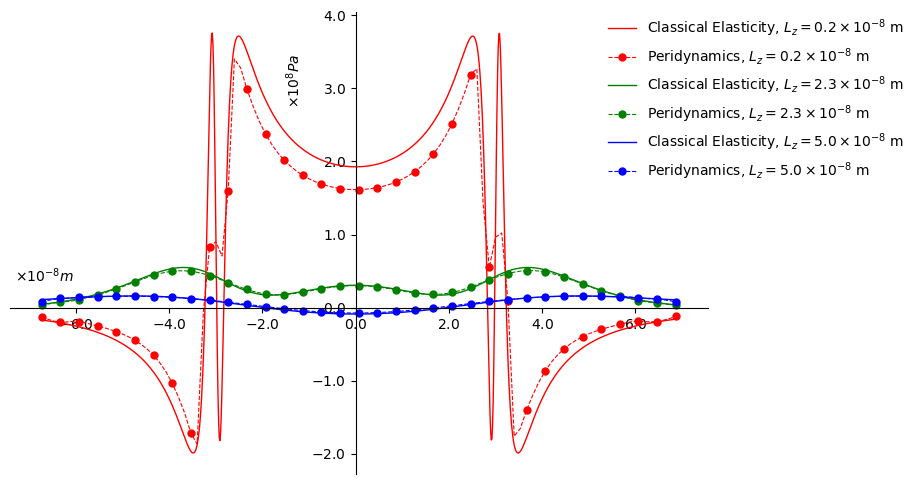}
			}
	\hspace{1pt}
\caption[]
{Comparision of the stress and displacement with classical elasticity solution. For all lines,  $L_y = 6.67 \times10^{-10} \si{m}$. Values on the horizontal axis of both subfigures refer to the X coordinates in Fig.\ref{fig:loop}.}
\label{fig:loopcompare}
\end{figure}

Besides straight dislocations mentioned above, the last case is a curved dislocation in an infinite domain. Fig.\ref{fig:loop} shows the geometry of the simulation domain, a cube with edge length $L=1.2\times10^{-7}\si{m}$.  A circular dislocation loop with radius $R=3\times10^{-8}\si{m}$ is placed in the $XY$ plane, and the Burgers vector is set as $\bm{b}=[b,0,0]$ with $b=2.5\times10^{-10} \si{m}$. The Young's modulus and the Poisson's ratio are $1.0\times10^{11} \si{Pa}$ and $0.34$, respectively. We apply the displacement solution in the classical elasticity to the boundaries. The classical elasticity solution is given in \citet{Hirth1983} and is numerically solved with adaptive integration. The horizon $\delta$ is $4.2\times10^{-9}\si{m}$, and the discretization parameters are $N=90$ and $M=3.15$.
\begin{figure}[!hbt]
	\centering
	\subfloat[][$u_{x}$]{
		\label{fig:loop_disp_x}
			\includegraphics[width=0.741\textwidth]{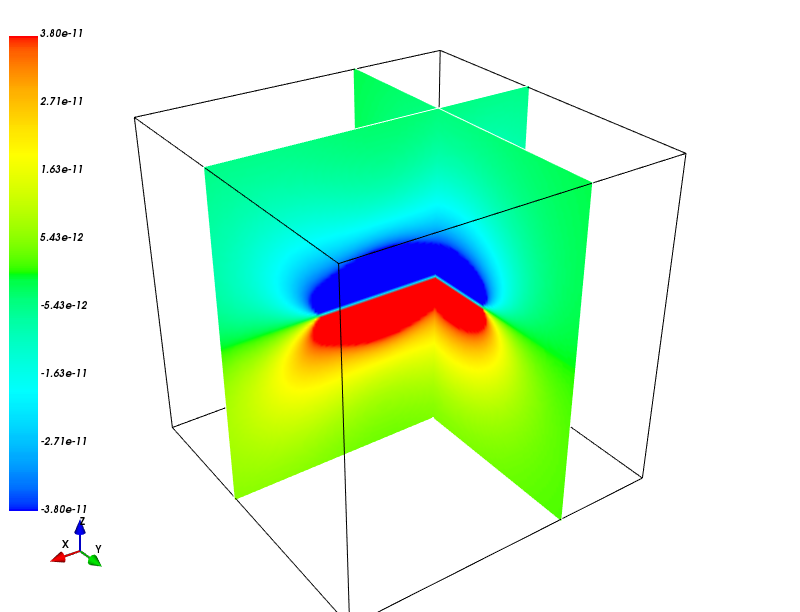}}\\
	\hspace{1pt}
	\subfloat[][$u_{y}$]{
		\label{fig:loop_disp_y}
			\includegraphics[width=0.741\textwidth]{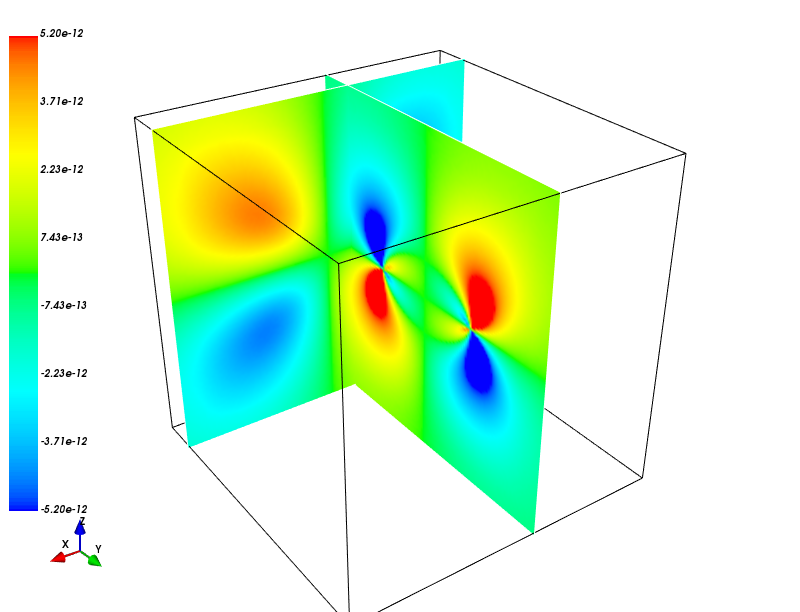}}
\caption[]
{The displacement field induced by a circular dislocation loop, unit: Pa. Slice position:\subref{fig:loop_disp_x}$x=2.11\times10^{-8} m$,$y=-2.31\times10^{-8} m$;\subref{fig:loop_disp_y}$x=1.15\times10^{-8} m$,$y=-4.61\times10^{-8} m$.}
\label{fig:loopdisp}
\end{figure}
\begin{figure}[htp]
	\centering
	\subfloat[][$\sigma_{xx}$]{
		\label{fig:loopwstressxx}
			\includegraphics[width=0.741\textwidth]{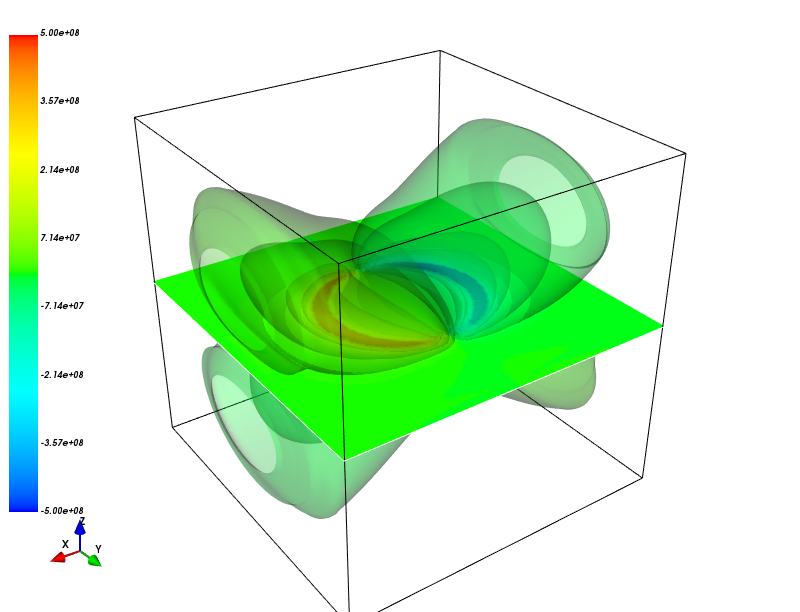}}\\
	\hspace{1pt}
	\subfloat[][$\sigma_{xy}$]{
		\label{fig:loopwstressxy}
			\includegraphics[width=0.741\textwidth]{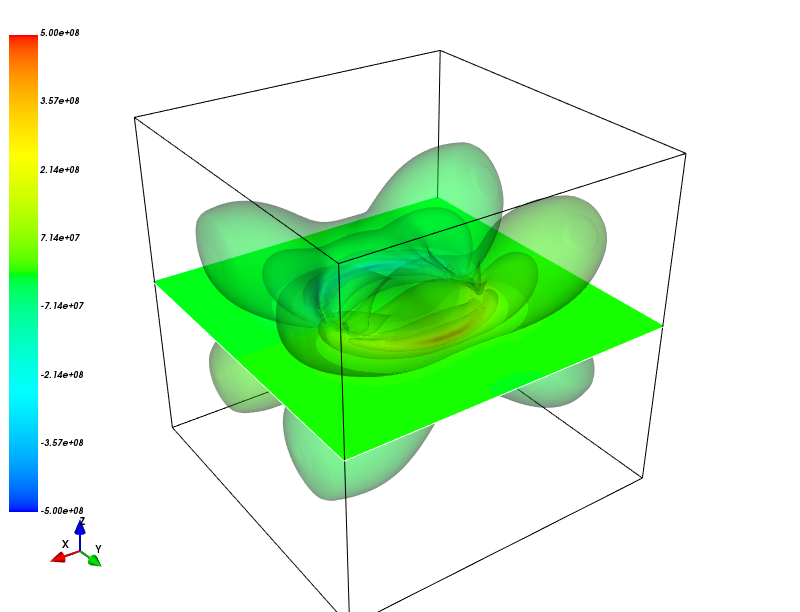}}
\caption[]
{3D contour and slice plane of the stress field induced by a circular dislocation loop, unit: Pa. Slice position: $z =4.6\times10^{-9}m$ }
\label{fig:loopstress}
\end{figure}
Fig.\ref{fig:loopdisp} shows the displacement induced by the circular dislocation loop. A displacement jump corresponding to the Burgers vector is clearly revealed in Fig.\ref{fig:loop_disp_x}. The stress field is presented in Fig.\ref{fig:loopstress}. No significant difference appears in both fields compared with literature results \citep{Khraishi2000}. Since the stress and displacement field created by circular dislocation is complex, a quantitive analysis is performed by plotting the numerical solution together with the classical elasticity solution along the line segment $(-L/2, L_y, L_z)$ to $(L/2, L_y, L_z)$, as shown in Fig.\ref{fig:loopcompare}. We sampled three line segments parallel to the X-axis by fixing $L_y$ and adjusting $L_z$. It can be seen from Fig.\ref{fig:loop_compare_dispz} that the displacement showing no difference between embedded discontinuity peridynamics and the classical elasticity, in line with the presented results of the screw dislocation. However, even though for line segments far from the glide plane the stress curves obtained by embedded discontinuity peridynamics still fit in well with the classical elasticity solution, differences exist along the line segment $L_z=0.2 \times 10^{-8} \si{m}$. The result may be explained by the nonsingular solution with embedded discontinuity, while another likely cause for the difference is the discretization. Similar errors also appear in XFEM modeling of dislocations \citep{Gracie2008}. The discretization utilizes $N=90$, a rough grid, by uniform node distribution, in which the loop curve is not considered, as shown in Fig.\ref{fig:volume}, neither the volume $V_\mathbf{X}$ is corrected. Consider the horizon is  $\delta=0.42 \times 10^{-8} \si{m}$, the occurrence of above numerical fluctuation could be explained as numerical errors in the modified Cauchy-Born rule. Therefore, the problem can be settled by refining the grid or decreasing the horizon size. Apart from the above numerical drawbacks, the result still shows the potential application value in modeling complex dislocations.
\section{Conclusions}
\label{sec:conclu}
In this paper, a nonlocal continuum framework of dislocations based on the state-based peridynamics has been constructed. Contrast to the previous practice of peridynamics in modeling of fracture-like discontinuity, the dislocation induced displacement discontinuity is embedded in the nonlocal constitutive model utilizing a modified Cauchy-Born rule.This approach extends the limits of the standard Cauchy-Born rule and avoids the surface effect which hinders the application of peridynamics. More broadly, the energy conservation between local and corresponding nonlocal continuum theories is enhanced for both intact and damaged media with evolving displacement discontinuity.
Compared with other dislocation models, the approach in this paper is capable of describing different types of dislocations without introducing additional parameters without clear physical meaning.  The introduction of nonlocality in peridynamics is via relaxing the strain energy density to a finite range via the horizon and the influence function. Clear meaning makes it possible for future applications of multiscale modeling. Though not included in this paper, fitting the influence function in the continuum is a promising way of modeling dislocation cores of different crystal lattices.
For verification, we examined different types of dislocations and the interaction between a pair of dislocations numerically. Surprisingly, singularities in the classical dislocation theories are regularized while only subtle distinction exists in the displacement field. We conclude the main findings as follows,
\begin{itemize}
  \item The concept of the Volterra dislocation can be perfectly reproduced with the embedded discontinuity peridynamics. As a pre-described displacement jump, the reproduced Burgers vector matches the defined one along the glide plane except the core region. For the near core region, the Burgers vector smoothly decreases to zero. 
  \item A benefit from the embedded treatment of interior discontinuities, surface effect is avoided for all cases. The interior surface effect disappears without any additional tracking or penalty.
  \item The stress solutions are nonsingular for both the edge and screw dislocations. The stress field near the dislocation core is in the same pattern as the nonsingular theory \citep{Cai2006} and the strain gradient solution \citep{Wang2016h}. As the decreasing of horizon, the embedded discontinuity peridynamics solution will converge to the classical elasticity solution.
  \item The displacement field computed with the embedded discontinuity peridynamics reaches an extremely high accuracy towards the classical elasticity. Though rough discretization was utilized for 3D cases, an accurate match between the classical elasticity and the nonlocal model is still reached. 
  \item The consistency of the driving force is guaranteed outside the core range, but the Peach-Koehler formula is not valid for computing dislocation driving force inside the dislocation core region. It indicates the mesh refinement is necessary for a low order discretization method and the particle arrangement in the present meshless method need more comprehension.
  \item Numerical instability occurs in the near core region for 3D curved dislocations. Apart from the non-singularity nature, it also indicates that the grid discretization in the particle based meshless method need careful rearrangement or refinement.
\end{itemize}

The work is the first step towards a multiscale dislocation dynamics framework. Currently, the Volterra type dislocation is modeled as a pre-described discontinuity, but the approach is opening doors for modeling unknown discontinuities such as complex fracture propagation modeling. Since the embedded discontinuity method is built upon the modified Cauchy-Born rule, the extension for complex constitutive modeling is straightforward. Though the current study is limited to linear elasticity, it is also possible to incorporate nonlinear elasticity for capturing the complex material behavior near dislocation cores. 


\bibliographystyle{plain}
\bibliography{library.bib}

\end{document}